\documentclass[journal]{IEEEtran}

\usepackage{amsmath}   
\usepackage{amssymb}   
\usepackage{bm}        
\usepackage{microtype} 
\usepackage{graphicx}  
\usepackage{tikz}      
\usepackage{booktabs}  
\usepackage{multirow}  
\usepackage{array}     
\usepackage{hyperref}  
\usepackage{cite}      
\usepackage{float}
\usepackage{enumitem}
\usepackage[table]{xcolor}
\definecolor{cyanblue}{RGB}{224,238,255}
\usepackage{graphicx}
\usepackage{subcaption}
\graphicspath{{figures/}} 
\usepackage{tcolorbox}
\usepackage[ruled,vlined]{algorithm2e}
\usepackage{booktabs}
\usepackage{siunitx}
\usepackage{longtable}
\usetikzlibrary{positioning,calc,arrows.meta,fit,shapes,backgrounds}

\usepackage{tikz}
\usepackage[edges]{forest}
\usepackage{xcolor}
\usetikzlibrary{calc}

\usepackage{adjustbox}
\usepackage{caption}
\renewcommand{\arraystretch}{1.1} 
\usepackage{amssymb}    
\usepackage{pifont}     
\usepackage{xcolor}     
 
\newcolumntype{Y}{S[table-format=1.3]}

\usepackage{bm}   
\usepackage{listings}
\usepackage{xcolor} 
\usepackage{listings}
\usepackage{xcolor}
\usepackage{xcolor}
\usepackage[T1]{fontenc}
\usepackage[utf8]{inputenc} 
\usepackage{listings}
\usepackage{xcolor}
\usepackage[T1]{fontenc}
\usepackage[utf8]{inputenc} 

\lstdefinelanguage{json}{
  morestring=[b]",
  moredelim=[s][\color{black}]{\{}{\}},
  moredelim=[s][\color{black}]{[}{]},
  stringstyle=\color{brown},
  showstringspaces=false,
}

\lstdefinestyle{jsonschema}{
  language=json,
  basicstyle=\ttfamily\footnotesize,
  frame=single,
  breaklines=true,
  columns=fullflexible,
  keepspaces=true,
  showstringspaces=false,
}
\lstdefinestyle{plain}{
  basicstyle=\ttfamily\footnotesize,
  frame=single,
  breaklines=true,
  columns=fullflexible,
  keepspaces=true,
  showstringspaces=false,
}

\lstdefinestyle{python}{
  language=Python,
  basicstyle=\ttfamily\footnotesize,
  frame=single,
  breaklines=true,
  columns=fullflexible,
  keepspaces=true,
  showstringspaces=false,
  keywordstyle=\color{blue},
  commentstyle=\color{gray},
  stringstyle=\color{brown},
}

\ifCLASSINFOpdf
\else
\fi
\begin{document}
%

\title{\textcolor{black}{$\alpha^3$-Bench: A Unified Benchmark of Safety, Robustness, and Efficiency for LLM-Based UAV Agents over 6G Networks}}

\author{
Mohamed~Amine~Ferrag$^{*\S}$,~\IEEEmembership{Senior~Member,~IEEE}, 
Abderrahmane~Lakas$^{*}$,~\IEEEmembership{Senior~Member,~IEEE},
and~Merouane~Debbah$^{1}$,~\IEEEmembership{Fellow,~IEEE}%
\thanks{$^{*}$Department of Computer and Network Engineering, College of Information Technology, United Arab Emirates University, Al Ain, United Arab Emirates.}%
\thanks{$^{1}$Khalifa University of Science and Technology, Abu Dhabi, United Arab Emirates.}%
\thanks{$^{\S}$Corresponding author: \texttt{mohamed.ferrag@uaeu.ac.ae}}%
}

%
%

\markboth{ }%
{Shell \MakeLowercase{\textit{et al.}}: Bare Demo of IEEEtran.cls for IEEE Journals}
%



\maketitle


\begin{abstract}
Large Language Models (LLMs) are increasingly being used as high-level controllers for autonomous Unmanned Aerial Vehicle (UAV) missions. However, current evaluations rarely assess whether such agents remain safe, protocol-compliant, and effective under realistic constraints of next-generation networking. This paper introduces $\alpha^{3}$-Bench, a benchmark for assessing LLM-driven UAV autonomy as a multi-turn conversational reasoning and control problem operating under dynamic 6G conditions. Each mission is modelled as a language-mediated control loop between an LLM-based UAV agent and a human operator, where decisions must satisfy strict schema validity, speaker alternation, mission policies, and safety bounds while adapting to fluctuating network slices (URLLC/eMBB/mMTC), latency, jitter, packet loss, throughput, and edge-load variations. To reflect modern agentic workflows, $\alpha^{3}$-Bench integrates a dual-action layer that supports both Model Context Protocol (MCP) tool calls and Agent-to-Agent (A2A) coordination, enabling the evaluation of tool-use consistency and multi-agent interactions within UAV missions. We construct a large-scale corpus of 113k AI conversational episodes grounded in UAVBench scenarios and evaluate 17 state-of-the-art LLMs using a fixed subset of 50 episodes per scenario under deterministic decoding. We propose a composite $\alpha^{3}$ metric that unifies six pillars—Task Outcome, Safety Policy, Tool Consistency, Interaction Quality, Network Robustness, and Communication Cost—with reliability- and efficiency-normalized scores per second and per thousand tokens. Experimental results show that several frontier models achieve near-perfect mission success and safety compliance (Task Outcome and Safety Policy $\geq$ 0.95), yet their robustness and efficiency diverge substantially: under degraded 6G conditions, Network Robustness scores drop by up to 30–40\%, while efficiency-normalized performance varies by more than a 2$\times$ factor across models in both $\alpha^{3}$ per-second and $\alpha^{3}$ per-1k-tokens metrics, reflecting significant variance in inference latency and token consumption. $\alpha^{3}$-Bench provides a reproducible, extensible foundation for benchmarking conversational UAV autonomy and guiding the development of safe, network-aware, and resource-efficient LLM agents in 6G-enabled aerial systems. To support open science and reproducibility, we release the $\alpha^{3}$-Bench dataset on GitHub: \url{https://github.com/maferrag/AlphaBench}.
\end{abstract}


%
\IEEEpeerreviewmaketitle

\section{Introduction} 

Recent advances in large language models (LLMs) have transformed autonomous decision-making systems by enabling agents to reason, plan, and interact through multi-turn natural-language dialog \cite{zhang2025nemotron,team2025kimi}. State-of-the-art LLMs now support long-context reasoning and complex tool use, positioning them as potential high-level controllers for real-world autonomous systems \cite{su2025toolorchestra,xiao2025uav}. At the same time, the global drone market is experiencing rapid expansion: one industry report estimates that the overall drone industry will grow from approximately USD 73.06 billion in 2024 to around USD 163.60 billion by 2030 at a compound annual growth rate (CAGR) of 14.3 \%, driven by improvements in autonomy, sensing, and operational capabilities. \cite{grandview2024drone} These developments motivate research on LLM-based autonomous agents for unmanned aerial vehicles (UAVs), where mission execution increasingly relies on conversation-mediated reasoning, structured interaction protocols, and adaptive decision-making under uncertainty \cite{sunderraman2024uav3d,yuan2025next}.

The challenge of autonomous UAV control is further amplified in the context of emerging 6G networks, which aim to succeed 5G technologies and support transformational applications requiring ultra-low latency, high reliability, and pervasive connectivity \cite{qu2025llm,jiang2025large}. 6G is currently being standardized as the next generation of cellular networks, with global research and regulatory activity targeting commercial deployments around 2030 \cite{ericsson6g}. Beyond speed and capacity improvements, 6G is expected to deliver orders-of-magnitude better performance (e.g., terabit-per-second data rates and sub-millisecond latencies) and integrate deeply with AI and edge computing to enable real-time orchestration of distributed autonomous systems \cite{sns2025whitepaper}. In this environment, effective UAV autonomy must account for time-varying network constraints—such as latency, jitter, throughput, and edge resource availability—which directly impact safety, coordination, and mission success \cite{abel2025large}.

From a standardization perspective, latency is a fundamental performance requirement in the evolution toward 6G networks. Building on the 5G framework, ongoing 3GPP studies for Release~19 and Release~20 explicitly envision further reductions in end-to-end latency toward sub-millisecond ranges to support future mission-critical services~\cite{3gpp_tr_22_870,3gpp_tr_38_914}. In the 3GPP 6G vision, service differentiation across ultra-reliable low-latency communication (URLLC), enhanced mobile broadband (eMBB), and massive machine-type communication (mMTC) remains a central design principle, with latency requirements tightly coupled to the communication role and service intent rather than the underlying platform~\cite{3gpp_tr_22_870}. Architecture evolution studies further emphasize that future 6G systems must support heterogeneous latency requirements within a unified framework through mechanisms such as network slicing and flexible service orchestration~\cite{3gpp_tr_23_700}. Together, these studies define a mission-centric latency landscape for 6G, in which latency is driven by service semantics and operational context rather than fixed per-slice assumptions.

Despite this convergence of advanced LLM capabilities, UAV autonomy growth, and 6G network evolution, existing evaluation frameworks remain limited \cite{ferrag2025llm}. Most benchmarks target static reasoning, tool-use robustness, or multimodal perception tasks, and they rarely model the continuous, interactive, safety-critical nature of UAV missions under fluctuating network conditions \cite{ferrag2025reasoning}. Moreover, current evaluations typically overlook the computational cost of reasoning, including latency and token usage, which are critical considerations for real-time, resource-constrained deployment. These gaps underscore the need for benchmarks that jointly assess reasoning quality, operational safety, network adaptivity, and efficiency, enabling rigorous comparison of LLM agents in scenarios that closely reflect the requirements of future 6G-enabled autonomous UAV systems \cite{hou2025model}.

Our study is guided by the following research questions, which aim to investigate how conversational reasoning, protocol-aware interaction, and 6G network dynamics shape the performance of LLM-based autonomous UAV agents:

\begin{tcolorbox}[
    colback=gray!10,
    colframe=black,
    arc=6pt,
    boxrule=0.7pt,
    left=2mm, right=2mm, top=1mm, bottom=1mm,
    title=Research Questions
]
\small
\begin{itemize}
    \item \textbf{RQ1:} How can multi-turn conversational interaction be formulated as a language-mediated control loop to evaluate reasoning, planning, and decision-making in autonomous UAV missions under realistic operational constraints?

    \item \textbf{RQ2:} To what extent does integrating structured action protocols—such as Model Context Protocol (MCP) \cite{modelcontextprotocol2025} tool calls and Agent-to-Agent (A2A) \cite{googleblog2025} communication—enable reliable assessment of tool consistency, protocol compliance, and multi-agent coordination in LLM-driven UAV systems?

    \item \textbf{RQ3:} How do dynamic 6G network conditions, including latency, packet loss, throughput variation, and edge-computing load, affect the reasoning strategies, safety awareness, and mission success of conversational UAV agents?

    \item \textbf{RQ4:} How do state-of-the-art LLMs differ in their ability to balance reasoning quality, safety policy adherence, and interaction coherence when evaluated across large-scale AI conversational UAV episodes derived from UAVBench scenarios \cite{ferrag2025uavbench} ?

    \item \textbf{RQ5:} What trade-offs emerge between reasoning performance, reliability, and computational efficiency when comparing LLMs using efficiency-normalized metrics such as $\alpha^{3}_{\text{per-sec}}$ and $\alpha^{3}_{\text{per-1k}}$ in 6G-enabled UAV missions?
\end{itemize}
\end{tcolorbox}

To address these research questions, we introduce $\alpha^{3}$-Bench, a large-scale benchmark for evaluating conversational reasoning, safety awareness, and network adaptability of LLM-based autonomous UAV agents operating under 6G communication conditions. $\alpha^{3}$-Bench formulates UAV mission execution as a multi-turn, language-mediated control problem, in which an LLM collaborates with a human operator through structured dialogue to plan, reason, and act in dynamically evolving environments. The benchmark is constructed on top of UAVBench scenarios \cite{ferrag2025uavbench} and extends them into AI conversational episodes represented as validated JSON dialogues that explicitly encode UAV states, airspace constraints, mission policies, and 6G network context.

$\alpha^{3}$-Bench integrates a dual-protocol interaction layer based on the Model Context Protocol (MCP) \cite{modelcontextprotocol2025} and Agent-to-Agent (A2A) \cite{googleblog2025} communication, enabling systematic evaluation of tool usage, protocol compliance, and multi-agent coordination within realistic UAV workflows. To ensure robustness and fairness at scale, the benchmark employs a controlled episode-generation pipeline with strict schema enforcement, multi-attempt recovery, deterministic seeding, and comprehensive failure accounting. In addition to reasoning quality, $\alpha^{3}$-Bench explicitly measures computational efficiency by tracking generation latency, token consumption, and provider-side usage, allowing joint assessment of effectiveness, reliability, and cost.

Built from 113k AI conversational episodes derived from UAVBench scenarios \cite{ferrag2025uavbench}, and evaluated using a fixed-budget subset of 50 episodes per model, $\alpha^{3}$-Bench provides a reproducible and extensible foundation for benchmarking LLM agents in safety-critical, network-constrained autonomous systems. The benchmark introduces a composite evaluation metric, $\alpha^{3}$, which aggregates task outcome, safety policy adherence, tool consistency, interaction quality, network robustness, and communication cost into a unified score. The key contributions of this work are summarized as follows:

\begin{itemize}
    \item We introduce $\alpha^{3}$-Bench, a novel benchmark that evaluates large language models as autonomous UAV agents through multi-turn conversational reasoning and control under dynamic 6G network conditions. Unlike prior benchmarks, it models UAV missions as language-mediated control loops rather than isolated reasoning or perception tasks.

    \item We propose a unified conversational decision framework that integrates UAV dynamics, airspace constraints, mission policies, and a 6G-aware network context. This formulation enables systematic evaluation of how LLMs adapt reasoning and control strategies in response to fluctuating latency, packet loss, and edge-computing conditions.

    \item We design a multi-protocol action layer that supports both Model Context Protocol (MCP) \cite{modelcontextprotocol2025} tool calls and Agent-to-Agent (A2A) \cite{googleblog2025} communication within each dialogue turn. This allows first-of-its-kind evaluation of protocol compliance, tool consistency, and multi-agent coordination in conversational UAV missions.

    \item We conduct an extensive experimental evaluation of \textbf{17 state-of-the-art LLMs}, spanning proprietary and open-weight families, using a large-scale corpus of 113k AI conversational UAV mission episodes grounded in UAVBench scenarios \cite{ferrag2025uavbench}. Each model is evaluated on a fixed subset of 50 episodes per scenario, ensuring fair, reproducible, and statistically robust comparison.

    \item We introduce a composite $\alpha^{3}$ metric that unifies six complementary pillars—Task Outcome, Safety Policy, Tool Consistency, Interaction Quality, Network Robustness, and Communication Cost—together with reliability- and efficiency-normalized scores. Our results reveal clear winners and losers, highlighting critical trade-offs between reasoning quality, safety, robustness to degraded 6G conditions, latency, and token efficiency.
\end{itemize}

The remainder of this paper is organized as follows. Section~\ref{sec:related_work} reviews existing benchmarks and evaluation frameworks for conversational agents, UAV autonomy, and embodied reasoning, and positions our work relative to prior approaches. Section~\ref{sec:problem_formulation} presents the formal problem formulation of $\alpha^3$-Bench, including the UAV environment model, conversational decision process, 6G network context, and the composite evaluation metrics. Section~\ref{sec:experimental_results} reports the experimental setup and comparative evaluation of 17 state-of-the-art LLMs, analyzing performance, reliability, and efficiency under realistic UAV mission conditions. Finally, Section~\ref{sec:conclusion} concludes the paper and discusses key findings, limitations, and directions for future research.

\begin{table*}[t]
\centering
\caption{Comparison of representative benchmarks related to conversational LLM agents and UAV autonomous decision-making.}
\label{tab:related_work_comparison}
\scriptsize
\setlength{\tabcolsep}{5pt}
\renewcommand{\arraystretch}{1.05}
\rowcolors{2}{white}{cyanblue!70} 
\begin{tabular}{lcccccc}
\hline
\textbf{Benchmark} 
& \textbf{AI Conversational} 
& \textbf{Tool / Agent Use} 
& \textbf{Multi-Agent} 
& \textbf{UAV-Specific} 
& \textbf{6G Network-Aware} 
& \textbf{Efficiency Metrics} \\
\hline
$\tau^2$-Bench~\cite{barres2025tau} 
& \checkmark 
& \checkmark 
& \checkmark 
& -- 
& -- 
& -- \\

AgentBench~\cite{liu2023agentbench}
& \checkmark
& \checkmark
& \checkmark
& -- 
& -- 
& -- \\

ACEBench~\cite{chen2025acebench} 
& \checkmark 
& \checkmark 
& \checkmark 
& -- 
& -- 
& -- \\

UAVBench~\cite{ferrag2025uavbench} 
& -- 
& -- 
& \checkmark 
& \checkmark 
& -- 
& -- \\

AirCopBench~\cite{zha2025aircopbench} 
& -- 
& -- 
& \checkmark 
& \checkmark 
& -- 
& -- \\

OpenUAV~\cite{wang2024towards} 
& -- 
& -- 
& -- 
& \checkmark 
& -- 
& -- \\

$\boldsymbol{\alpha^3}$-\textbf{Bench (Ours)} 
& \checkmark 
& \checkmark
& \checkmark 
& \checkmark 
& \checkmark 
& \checkmark \\
\hline
\end{tabular}
\end{table*}

\section{Related Work}\label{sec:related_work}

Recent research on large language models has produced a diverse set of benchmarks aimed at evaluating conversational reasoning, tool use, multimodal perception, and embodied decision-making. These efforts span abstract conversational environments, general-purpose agentic workflows, and domain-specific settings such as autonomous aerial systems. However, existing benchmarks differ substantially in their assumptions, interaction models, and evaluation objectives, making direct comparison challenging. In this section, we organize prior work into three main categories: (i) conversational and tool-use benchmarks for LLM agents, (ii) UAV-specific benchmarks targeting reasoning, perception, and decision-making, and (iii) embodied vision--language navigation frameworks for aerial platforms. This structured review highlights the strengths and limitations of existing approaches and clarifies the unique design choices that motivate the development of $\alpha^3$-Bench. Table~\ref{tab:related_work_comparison} presents a comparative overview of representative benchmarks related to conversational large language model agents and autonomous UAV decision-making. The table contrasts prior work along key dimensions, including conversational interaction, tool and agent usage, multi-agent coordination, UAV-specific task modeling, network awareness, and efficiency-oriented evaluation. The comparison highlights that existing benchmarks typically address only a subset of these aspects, whereas $\alpha^3$-Bench uniquely integrates conversational autonomy, structured tool and agent interactions, network-aware reasoning under 6G conditions, and efficiency metrics within a unified UAV-focused evaluation framework.

\subsection{Conversational and Tool-Use Benchmarks for LLM Agents}

Barres et al. \cite{barres2025tau} introduce the $\tau^2$-Bench, a benchmark for assessing conversational agents in settings where both the agent and the user can actively modify the environment through tool interactions. This framework moves beyond earlier single-control benchmarks by formulating the problem as a Dec-POMDP, allowing both parties to observe, act, and communicate within a shared state. The benchmark provides a telecom troubleshooting domain and a programmatic task-generation pipeline that constructs diverse, verifiable scenarios from smaller building blocks. In addition, it incorporates a constrained user simulator whose behavior is shaped by the available tools and environmental state, enabling more reliable user-side interactions. Experimental results show that current language agents struggle significantly when required to collaborate with an active user, highlighting the importance of dual-control conversational evaluation.

Chen et al. \cite{chen2025acebench} propose ACEBench, a comprehensive benchmark designed to evaluate how large language models employ tools across a wide spectrum of scenarios, ranging from simple single-turn interactions to complex multi-turn, multi-step environments. ACEBench introduces three data categories—Normal, Special, and Agent—targeting different aspects of tool use, including basic function invocation, handling incomplete or erroneous instructions, and coordinated actions within sandboxed real-world simulations. By combining synthetic API generation, multi-agent dialogue construction, and expert-curated scenarios, the benchmark provides a structured framework for examining robustness and failure modes in tool-augmented reasoning.

Liu et al.~\cite{liu2023agentbench} introduce AGENTBENCH, a multi-dimensional benchmark designed to evaluate large language models as autonomous agents across eight interactive environments, including operating systems, databases, knowledge graphs, games, and web-based tasks. AGENTBENCH formalizes LLM--agent interaction as a partially observable Markov decision process (POMDP) and evaluates agents on multi-round instruction following, reasoning, and action execution using real executable environments. By covering code-grounded, game-grounded, and web-grounded scenarios, AGENTBENCH provides one of the first systematic evaluations of LLMs as general-purpose agents operating beyond static question-answering settings.

Despite its breadth, AGENTBENCH primarily targets abstract or desktop-scale interaction domains and does not model embodied autonomy, domain-specific mission constraints, or communication-system dynamics. In contrast to $\alpha^3$-Bench, AGENTBENCH does not consider UAV-specific operational contexts, network-induced uncertainty, or adaptive behavior under constrained communication regimes. Moreover, agent actions in AGENTBENCH are evaluated independently of latency, reliability, or resource-efficiency considerations, which are central to safety-critical autonomous aerial missions.

\subsection{Benchmarks for UAV Reasoning and Autonomous Decision-Making}

Ferrag et al. \cite{ferrag2025uavbench} propose UAVBench, a large-scale benchmark for evaluating the reasoning, perception, and decision-making capabilities of LLM-driven autonomous UAV systems under realistic mission conditions. The benchmark introduces 50{,}000 LLM-generated and safety-validated flight scenarios encoded in a structured JSON format, capturing mission objectives, UAV configurations, environmental factors, and quantitative risk assessments. The authors further extend this effort with UAVBench\_MCQ, a companion benchmark of 50{,}000 multiple-choice questions spanning diverse reasoning styles such as aerodynamics, navigation, ethics, and multi-agent coordination. Their evaluation of 32 state-of-the-art models reveals strong performance in policy and perception reasoning, but persistent weaknesses in ethically constrained and resource-limited decision-making.

Zha et al. \cite{zha2025aircopbench} introduce AirCopBench, a benchmark for evaluating multimodal large language models in multi-UAV collaborative perception and embodied reasoning under challenging real-world conditions. The dataset integrates 2.9k multi-view aerial images and 14.6k VQA pairs across 14 task types, covering scene understanding, object reasoning, perception assessment, and collaborative decision-making. AirCopBench further incorporates realistic perception degradations such as occlusion, blur, noise, and data loss, and relies on a structured pipeline combining simulation, real UAV footage, and human annotation. Experimental results demonstrate substantial limitations in multi-view reasoning and collaborative tasks, with top models performing far below human baselines.

\subsection{Vision--Language Navigation and Embodied UAV Benchmarks}

Wang et al. \cite{wang2024towards} propose a unified framework for realistic UAV vision--language navigation, introducing OpenUAV, a high-fidelity Unreal Engine–based simulation platform supporting continuous 6-DoF flight control, multi-sensor payloads, and realistic physics. Using this platform, the authors construct a large-scale UAV VLN dataset comprising over 12k human-flown trajectories, each paired with detailed natural-language instructions grounded in complex outdoor environments. To address limitations of conventional navigation tasks, they further introduce UAV-Need-Help, an assistant-guided object search benchmark that evaluates UAV performance under varying levels of real-time guidance. Their results demonstrate notable improvements over classical VLN baselines, while revealing a significant performance gap relative to human operators.

\subsection{Positioning of $\alpha^3$-Bench}

Unlike prior benchmarks that primarily focus on generic tool use (e.g., ACEBench) or dual-control conversational coordination in abstract domains (e.g., $\tau^2$-Bench), $\alpha^3$-Bench is specifically designed to evaluate \emph{end-to-end conversational autonomy} for UAV missions operating under \emph{dynamic 6G communication constraints}. In our formulation, each episode constitutes a language-mediated control loop in which an LLM acts as an autonomous UAV controller, collaborating with a human operator over multiple dialogue turns while continuously conditioning on a 6G network state vector encompassing slice selection, latency, jitter, packet loss, throughput, and edge-compute load.

This design fundamentally distinguishes $\alpha^3$-Bench from UAV-focused datasets such as UAVBench , AirCopBench, and OpenUAV, which emphasize scenario construction, perception, or embodied navigation, but do not jointly operationalize conversational decision-making, network-aware adaptation, and resource efficiency within a unified evaluation framework. Moreover, $\alpha^3$-Bench explicitly models modern agentic workflows by embedding a multi-protocol action layer into each dialogue turn, supporting both Model Context Protocol (MCP) \cite{modelcontextprotocol2025} tool calls and Agent-to-Agent (A2A) \cite{googleblog2025} coordination. Finally, beyond task outcome and safety, $\alpha^3$-Bench introduces a composite evaluation metric integrating six complementary pillars and reports efficiency-normalized scores, enabling systematic comparison of LLM agents not only by what they achieve, but also by how reliably and efficiently they achieve it under realistic 6G-enabled UAV operating conditions.

\begin{figure*}[t]
    \centering
    \includegraphics[width=\linewidth]{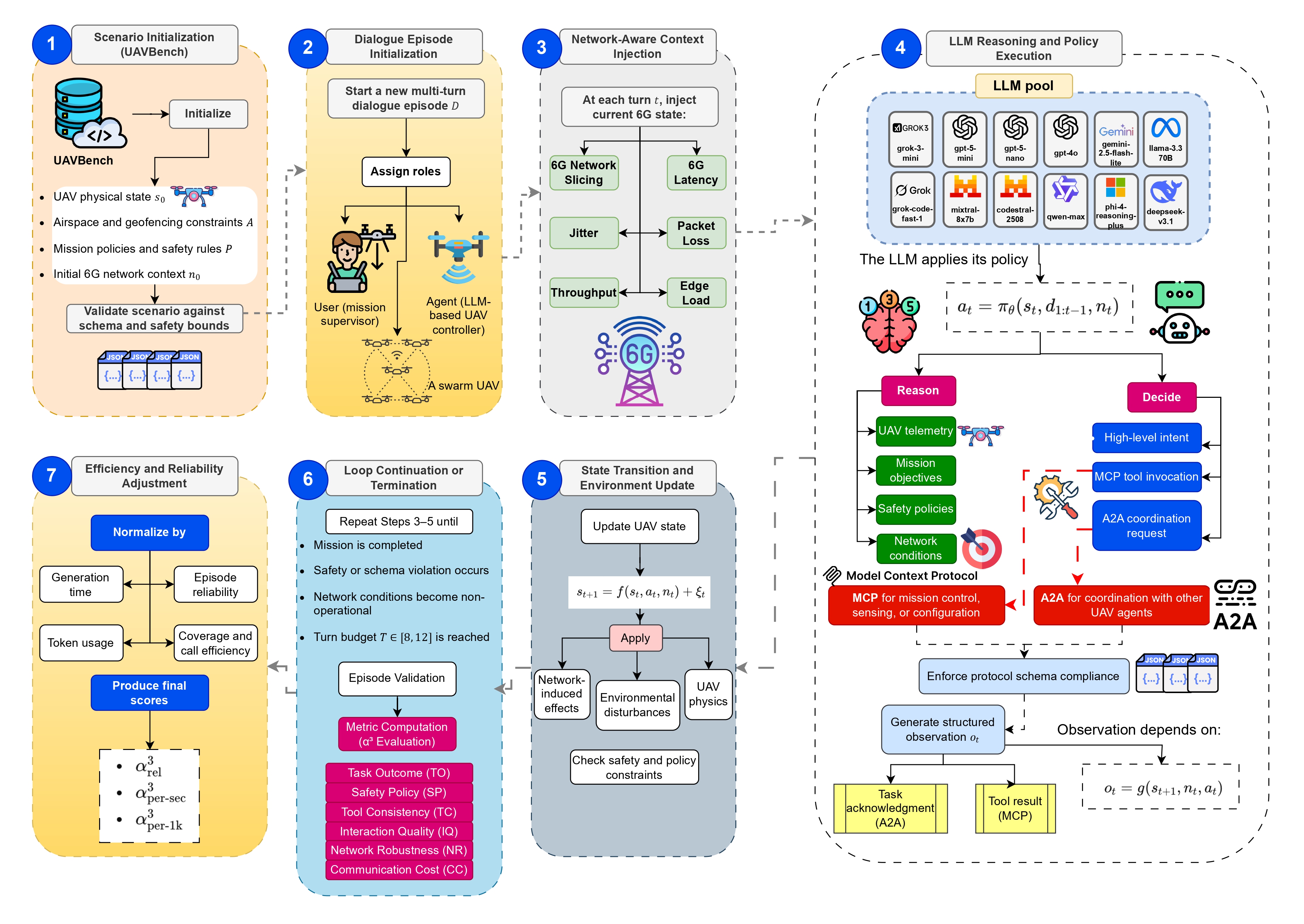}
   \caption{End-to-end workflow of the $\alpha^3$-Bench framework for evaluating LLM-based UAV agents under dynamic 6G communication conditions. The figure illustrates scenario initialization from UAVBench, dialogue-based mission execution with network-aware reasoning, structured action invocation via MCP and A2A protocols, environment and state updates, loop termination, and final efficiency- and reliability-adjusted $\alpha^3$ evaluation metrics.}
    \label{fig:fig2}
\end{figure*}

\section{Problem Formulation}\label{sec:problem_formulation}

The objective of this work is to evaluate the reasoning capacity, safety awareness, and network adaptability of Large Language Models (LLMs) within autonomous Unmanned Aerial vehicles (UAV) missions operating under 6G communication conditions. 
The problem is formulated as a multi-turn conversational reasoning and control task, where the LLM assumes the role of an autonomous UAV agent collaborating with a human operator to plan, execute, and complete flight missions safely and efficiently. To ensure a fair cross-model comparison, $\alpha^3$-Bench also measures the token
consumption, computational latency, and provider-side cost for every generated
episode, enabling a joint evaluation of reasoning quality and resource efficiency.

Fig. \ref{fig:fig2} presents the overall workflow of $\alpha^3$-Bench, which formalizes UAV mission execution as a language-mediated, multi-turn conversational control process operating under realistic 6G network conditions. The workflow begins with scenario initialization from UAVBench, where the UAV physical state, airspace and geofencing constraints, mission policies, and initial 6G network context are instantiated and validated. A structured dialogue episode is then initialized between a human supervisor and an LLM-based UAV agent. At each dialogue turn, the current 6G network state, including slicing, latency, jitter, packet loss, throughput, and edge load, is injected as contextual input to condition the agent’s reasoning. The LLM applies its policy to jointly reason over UAV telemetry, mission objectives, safety constraints, and network conditions, producing either a high-level intent or a structured action through the Model Context Protocol (MCP) \cite{modelcontextprotocol2025} or Agent-to-Agent (A2A) \cite{googleblog2025} coordination. The selected action triggers a corresponding observation, after which the UAV state evolves according to physical dynamics, network-induced effects, and environmental disturbances, while continuously enforcing safety and policy constraints. This closed-loop process iterates until mission completion, constraint violation, degraded network termination, or exhaustion of the turn budget. Finally, the generated episode is validated and evaluated using the composite $\alpha^3$ metric, integrating task outcome, safety compliance, protocol consistency, interaction quality, network robustness, and communication cost, followed by efficiency- and reliability-based normalization to enable fair cross-model comparison.

\subsection{Environment Modeling}

The Unmanned Aerial Vehicle (UAV) operational environment is modelled as a multi-domain system controlling flight dynamics, airspace constraints, network conditions, and mission policies. 
Formally, the environment is defined as
\begin{equation}
\mathcal{E} = (\mathcal{U}, \mathcal{A}, \mathcal{N}, \mathcal{S}, \mathcal{P}),
\end{equation}
where:
\begin{itemize}
    \item $\mathcal{U}$ represents the UAV dynamics and sensory configuration, including motion models and onboard sensing capabilities;
    \item $\mathcal{A}$ denotes the spatial airspace representation and geofencing constraints that restrict UAV movement;
    \item $\mathcal{N}$ characterizes the sixth-generation (6G) communication network layer and its associated Quality-of-Service (QoS) attributes, such as latency, throughput, and reliability;
    \item $\mathcal{S}$ is the continuous UAV state space, capturing the physical and operational variables of the platform;
    \item $\mathcal{P}$ defines the mission policies and safety constraints that govern permissible UAV actions during flight execution.
\end{itemize}

\subsubsection{UAV Dynamics and Sensors}
The UAV dynamics encapsulate translational and rotational motion, represented by a state vector:

\begin{equation}
s_t = (x_t, y_t, z_t, v_t, \psi_t, \text{battery}_t),
\end{equation}

where $(x_t, y_t, z_t)$ denotes the position of the UAV in Cartesian coordinates, $v_t$ its velocity magnitude, $\psi_t$ its direction (yaw angle), and $\text{battery}_t \in [0,100]$ the remaining percentage of energy.

The UAV perceives its environment through a multimodal sensor set $\mathcal{U}_{\text{sense}} = \{\text{LiDAR}, \text{RGB}, \text{Thermal}, \text{IMU}\}$.

Each sensor $s_i \in \mathcal{U}_{\text{sense}}$ contributes to the observation model $o_t$ used by the reasoning agent to infer the next intent or action.

\subsubsection{Airspace and Geofencing Constraints}

The airspace domain $\mathcal{A}$ defines the operational altitude limits and the set of no-fly zones (NFZs) that restrict UAV motion. 
Let $z_{\min}$ and $z_{\max}$ denote the minimum and maximum allowable flight altitudes, and let $\mathcal{G}$ represent the set of geofenced regions:
\begin{equation}
\mathcal{A} = \{ (z_{\min}, z_{\max}), \mathcal{G} \}, 
\quad 
\mathcal{G} = \{ g_i = (C_i, r_i) \},
\end{equation}
where each geofence $g_i$ represents a cylindrical or polygonal exclusion zone, defined by its center $C_i$ and radius $r_i$ (or an equivalent polygonal boundary).

A safety constraint is enforced such that the UAV trajectory must satisfy:
\begin{equation}
z_{\min} \leq z_t \leq z_{\max}, \quad \forall t,
\qquad \text{and} \qquad
d(p_t, C_i) \geq r_i, \; \forall g_i \in \mathcal{G},
\end{equation}
where $p_t = (x_t, y_t, z_t)$ denotes the UAV position at time $t$, and $d(\cdot)$ represents the Euclidean distance metric.

\subsubsection{6G Network Context}

The sixth-generation (6G) network environment $\mathcal{N}$ provides the communication and edge-computing backbone for the control of the Unmanned Aerial Vehicle (UAV) mission. 
At each time step $t$, the network state is defined as:
\begin{equation}
n_t = (\text{slice}_t, \text{lat}_t, \text{jit}_t, \text{loss}_t, \text{thr}_t, \text{edge}_t),
\end{equation}
where:
\begin{itemize}
    \item $\text{slice}_t \in $ Ultra- Reliable Low-Latency Communications (URLLC),
    enhanced Mobile Broadband (eMBB), massive Machine-Type Communications (mMTC) identifies the active 6G service type;
    \item $\text{lat}_t$ and $\text{jit}_t$ denote end-to-end communication latency and jitter, measured in milliseconds;
    \item $\text{loss}_t$ represents the packet loss rate expressed as a percentage;
    \item $\text{thr}_t$ denotes the achievable network throughput in megabits per second (Mbps);
    \item $\text{edge}_t \in [0,1]$ models the normalized computational load of the edge server supporting the UAV.
\end{itemize}

These network parameters dynamically evolve during the dialogue, allowing the Large Language Model (LLM)-based controller to adapt mission reasoning and communication strategies to degraded or fluctuating connectivity conditions, which is a critical requirement for 6G-enabled UAV operations.

\subsubsection{Mission Policies and Safety Bounds}

Mission policies $\mathcal{P}$ formalize the operational and ethical constraints that the Unmanned Aerial Vehicle (UAV) must follow during mission execution. 
These policies include altitude envelopes, minimum energy thresholds, collision avoidance margins, and communication reliability requirements.

The policy layer acts as a supervisory constraint on the action space $\mathcal{A}_t$ of the Large Language Model (LLM)-based agent, ensuring that:
\begin{equation}
a_t \in \mathcal{A}_{\text{safe}}(s_t, \mathcal{P}) \subseteq \mathcal{A}_{\text{total}}.
\end{equation}
This mechanism prevents unsafe or infeasible behaviors while preserving flexibility in high-level, mission-oriented decision-making.

\subsubsection{State Evolution and Kinematics}

The initial state $\mathbf{s}_0$ is sampled from UAVBench scenarios [18] and evolves according to the discrete-time transition law:
\begin{equation}
    {s}_{t+1} = f({k}_t, {a}_t, n_t) + {\xi}_t,
    \label{eq:state_evolution}
\end{equation}
where $f$ encapsulates the coupled UAV kinematics and dynamics alongside network feedback $n_t$. Specifically, for a kinematic state vector comprising position $\mathbf{p}_t \in \mathbb{R}^3$ and velocity $\mathbf{v}_t \in \mathbb{R}^3$, the kinematic formulation within $f(\cdot)$ follows the discretization:
\begin{equation}
    {k}_t = \begin{bmatrix} \mathbf{p}_t \\ \mathbf{v}_t \\ \boldsymbol{\Theta}_t \\ \boldsymbol{\omega}_t \end{bmatrix},
    \label{eq:state_def}
\end{equation}
where $\mathbf{p}_t \in \mathbb{R}^3$ is the inertial position, $\mathbf{v}_t \in \mathbb{R}^3$ is the linear velocity, $\boldsymbol{\Theta}_t \in \mathbb{R}^3$ represents the Euler angles (attitude), $\boldsymbol{\omega}_t \in \mathbb{R}^3$ denotes angular rates, and such that:

\begin{equation}
    \begin{aligned}
        \mathbf{p}_{t+1} &= \mathbf{p}_t + \mathbf{v}_t \Delta t + \frac{1}{2} \mathbf{R}(\boldsymbol{\Theta}_t) \frac{\mathbf{T}_t}{m} \Delta t^2, \\
        \mathbf{v}_{t+1} &= \mathbf{v}_t + \left( \mathbf{R}(\boldsymbol{\Theta}_t) \frac{\mathbf{T}_t}{m} - \mathbf{g} \right) \Delta t,
    \end{aligned}
\end{equation}
where $\mathbf{R}(\boldsymbol{\Theta}_t)$ is the rotation matrix derived from the attitude state, $\mathbf{T}_t$ is the thrust vector, $m$ is the mass, and $\mathbf{g}$ is the gravity vector.

The term ${\xi}_t$ represents the stochastic disturbance capturing environmental uncertainty (e.g., wind turbulence and sensor noise). We model ${\xi}_t$ as a multivariate Gaussian process:
\begin{equation}
    {\xi}_t \sim \mathcal{N}(\mathbf{0}, \boldsymbol{\Sigma}_{\xi}),
\end{equation}
where ${\Sigma}_{\xi}$ is the covariance matrix characterizing the intensity of the disturbances. The joint environment $\mathcal{E}$ therefore couples these motion dynamics with network constraints and mission policies, creating a unified testbed for evaluating the reasoning, planning, and adaptability capabilities of Large Language Model (LLM)-based UAV controllers under realistic sixth-generation (6G) communication conditions.


\subsection{Conversational Decision Process}

The UAV mission is formulated as a multi-turn dialogue between two entities: the \textit{agent}, which is an LLM-based UAV controller, and the \textit{user}, representing a human operator or mission supervisor.
Each dialogue turn corresponds to a reasoning and control iteration, during which the LLM interprets the mission context, reasons over the current UAV and network state, and proposes the next high-level action conditioned on the prevailing 6G network conditions.

\begin{table*}[ht]
\centering
\caption{Top-10 most frequently expressed high-level agent intents across more than 113k conversational UAV mission episodes.}
\label{tab:intent_top10}
\scriptsize
\setlength{\tabcolsep}{5pt}
\renewcommand{\arraystretch}{1.05}
\rowcolors{2}{white}{cyanblue!70} 
\begin{tabular}{lccc}
\toprule
Intent & Count & Share of Intents (\%) & Avg. per Episode \\
\midrule
initiate mission and check telemetry & 30{,}630 & 2.53 & 0.27 \\
initiate\_mission                  & 23{,}986 & 1.98 & 0.21 \\
confirm\_mission\_start            & 21{,}592 & 1.78 & 0.19 \\
initiate\_survey\_mission          & 16{,}810 & 1.39 & 0.15 \\
request\_swarm\_coordination       & 10{,}956 & 0.91 & 0.10 \\
adapt\_to\_network\_degradation    & 10{,}771 & 0.89 & 0.09 \\
detect\_network\_degradation       & 10{,}433 & 0.86 & 0.09 \\
detect network degradation and adapt & 9{,}372 & 0.77 & 0.08 \\
return\_to\_base                   & 8{,}629  & 0.71 & 0.08 \\
request\_thermal\_scan             & 7{,}177  & 0.59 & 0.06 \\
\bottomrule
\end{tabular}
\end{table*}

\subsubsection{Dialogue Representation}

Each mission is represented as a structured sequence of dialogue turns:
\begin{equation}
\mathcal{D} = \{ d_t = (r_t, a_t, o_t, n_t) \mid t = 1,\dots,T \}.
\end{equation}

Each dialogue turn $d_t$ consists of the following components:
\begin{itemize}
    \item $r_t \in \{\text{agent},\text{user}\}$ denotes the speaker role, where the agent corresponds to the Large Language Model (LLM)-based UAV controller and the user represents the human mission supervisor.

    \item $a_t$ denotes the high-level intent, optionally accompanied by a structured action. Specifically,
    \begin{equation}
    a_t \in 
    \left\{
    \begin{aligned}
        &(\text{ MCP protocol},\;\emph{name},\;\emph{args}), \\
        &(\text{ A2A protocol},\;\emph{task},\;\emph{to},\;\emph{payload}), \\
        &\text{\textit{intent-only}}
    \end{aligned}
    \right\},
    \end{equation}
    where the \emph{intent-only} case represents a high-level mission decision that does not invoke an explicit protocol action.

    \item $o_t$ denotes the structured observation generated in response to the action:
    \begin{equation}
    o_t \in 
    \left\{
    \begin{aligned}
        &(\emph{tool},\;\emph{result}), \\
        &(\emph{task},\;\emph{from},\;\emph{status},\;\emph{payload})
    \end{aligned}
    \right\}.
    \end{equation}

    \item $n_t$ represents the current state vector of the sixth-generation (6G) communication network.
\end{itemize}

To preserve a human-in-the-loop interaction model, strict alternation between speaker roles is enforced:
\begin{equation}
r_t \neq r_{t-1}, \quad \forall t \ge 2.
\end{equation}

\begin{figure*}[t]
    \centering
    \includegraphics[width=\linewidth]{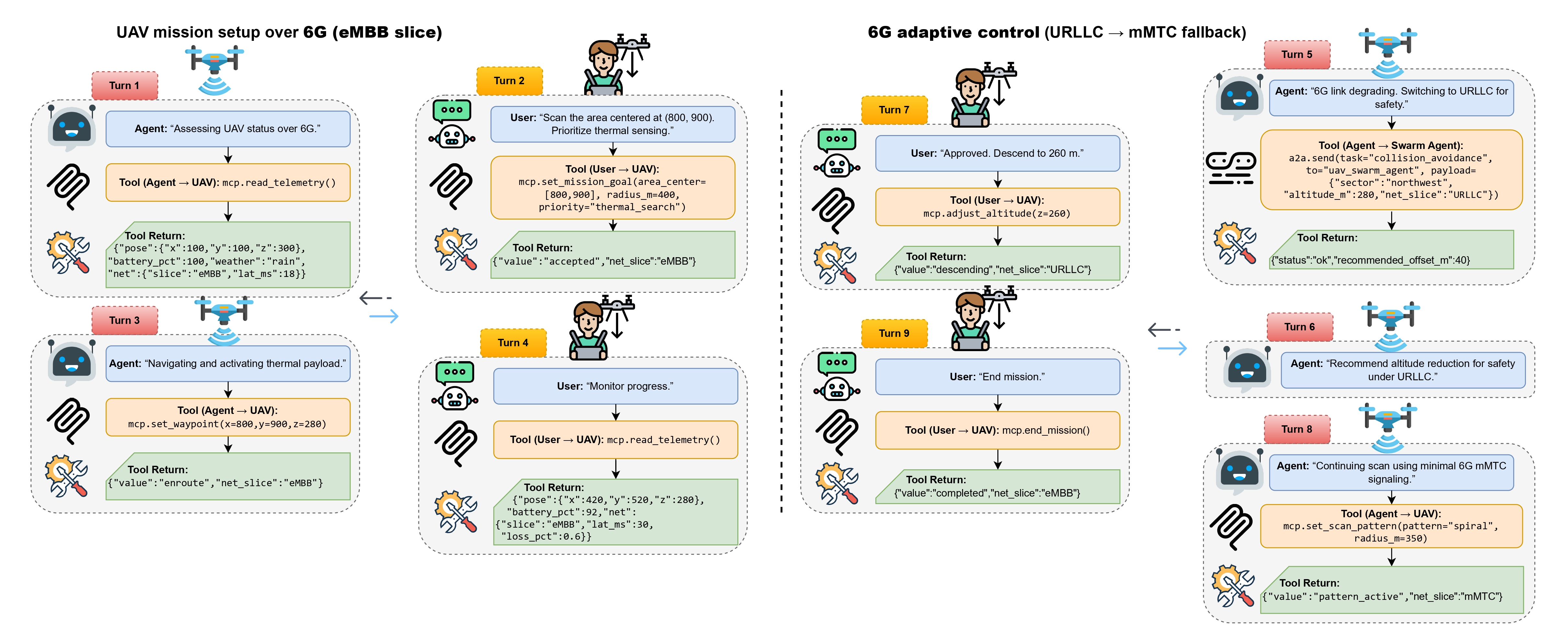}
   \caption{An example agent--user interaction trajectory in the UAV domain of $\alpha^3$-Bench under 6G communication. 
The left panel illustrates user--agent interactions via the Model Context Protocol (MCP) \cite{modelcontextprotocol2025}, where the UAV state is queried and a thermal area-scan mission is initiated over a 6G eMBB slice. 
The right panel highlights the agent’s adaptive decision-making under dynamic network conditions, including coordination with other autonomous agents through the agent-to-agent (A2A) \cite{googleblog2025} protocol for collision avoidance, and seamless switching between eMBB, URLLC, and mMTC 6G network slices to preserve safety and mission continuity. }
    \label{fig:fig1}
\end{figure*}

Table~\ref{tab:intent_top10} presents the ten most frequently expressed high-level agent intents aggregated over more than 113k conversational UAV mission episodes in $\alpha^3$-Bench. The results reveal a strong emphasis on mission initialization and state validation, with intents such as \texttt{initiate mission and check telemetry}, \texttt{initiate\_mission}, and \texttt{confirm\_mission\_start} collectively dominating the distribution. This indicates that LLM-based UAV agents consistently ground their reasoning in explicit mission setup and early situational awareness before committing to execution. In particular, several of the most frequent intents are explicitly network-aware, including \texttt{detect\_network\_degradation}, \texttt{adapt\_to\_network\_degradation}, and their combined variants, highlighting that communication conditionexplicitly considered at the intent level rather than being treated as incidental constraints. Intents related to multi-agent collaboration, such as \texttt{request\_swarm\_coordination}, and safety-oriented termination behaviors, such as \texttt{return\_to\_base}, further demonstrate that conversational UAV control in $\alpha^3$-Bench is governed by coherent, goal-directed reasoning cycles that integrate mission objectives, network conditions, and coordination requirements.

\subsubsection{User Speaker Modeling}

\textcolor{black}{
In $\alpha^3$-Bench, the \emph{user} speaker is always simulated and does not correspond to a real human participant. The user role is introduced to represent a mission supervisor or operator interacting with the LLM-based UAV agent within a language-mediated control loop. This design preserves a human-in-the-loop interaction structure while enabling fully automated, large-scale, and reproducible evaluation. User turns are generated using three complementary mechanisms, selected according to the benchmark split and evaluation objective. First, \emph{scripted prompt templates} are used for baseline scenarios to ensure deterministic task specification and controlled coverage of canonical UAV missions (e.g., navigation, inspection, survey, delivery, and recovery). These templates follow fixed linguistic patterns with parameterized fields derived directly from the UAVBench scenario state, such as target coordinates, altitude bounds, sensing requirements, and mission constraints. Second, \emph{LLM-generated user prompts} are employed in complex or adaptive scenarios to simulate realistic mission evolution under dynamic conditions. In this setting, a separate language model generates user instructions conditioned on the current dialogue history, UAV state, mission progress, and 6G network context, while remaining constrained by predefined safety and policy rules. This mechanism introduces linguistic variability, contextual dependencies, and multi-step supervisory behavior without violating structural or safety constraints. Third, \emph{fixed user prompts} are used for all evaluation splits to guarantee full reproducibility and fair comparison across models. These prompts are frozen and shared as part of the benchmark corpus, ensuring that every evaluated LLM receives identical user inputs for a given episode. Across all settings, strict alternation between agent and user roles is enforced, and user messages are restricted to high-level mission supervision rather than direct low-level control, ensuring consistency with the conversational UAV control formulation of $\alpha^3$-Bench.
}

\subsubsection{Multi-Protocol Action Layer (MCP and A2A)}

To accurately model modern UAV workflows and emerging standards in agent communication, 
$\alpha^{3}$-Bench integrates two structured protocols into each dialog turn, namely the Model Context Protocol (MCP) \cite{modelcontextprotocol2025} and the Agent-to-Agent Protocol (A2A) \cite{googleblog2025}. Figure \ref{fig:fig1} presents an illustrative interaction trajectory from the UAV domain of $\alpha^3$-Bench, highlighting how agents based on the large language model (LLM) operate under realistic 6G communication constraints. The scenario demonstrates a multi-turn mission in which the agent and a human operator collaboratively plan, execute, and monitor a thermal area-scan task using structured mission control protocol (MCP) \cite{modelcontextprotocol2025} tool calls. Throughout the mission, the agent reasons jointly over UAV telemetry, payload constraints, airspace policies, and network conditions, dynamically adapting its behavior as link quality degrades. To ensure safety and mission continuity, the agent coordinates with autonomous peer entities via the agent-to-agent (A2A) \cite{googleblog2025} protocol to avoid collisions, while seamlessly switching between eMBB, URLLC, and mMTC 6G network slices.

\paragraph{Model Context Protocol (MCP) \cite{modelcontextprotocol2025}}

The Model Context Protocol (MCP) defines structured interactions between the Large Language Model (LLM)-based UAV controller and mission-level tools, such as sensing, navigation, or configuration services. 
An MCP action is represented as:
\begin{equation}
a_t = (\text{mcp},\;\text{name},\;\text{args}),
\end{equation}
and must generate a corresponding observation of the form:
\begin{equation}
o_t = (\text{tool}=\text{name},\;\text{result}),
\end{equation}
where \texttt{name} identifies the invoked tool and \texttt{args} specifies its input parameters.

\paragraph{Agent-to-Agent Protocol (A2A) \cite{googleblog2025}}

The Agent-to-Agent Protocol (A2A) models structured interactions between the UAV and other autonomous agents or subsystems operating in the same environment. 
An A2A action is represented as:
\begin{equation}
a_t = (\text{a2a},\;\text{task},\;\text{to},\;\text{payload}),
\end{equation}
with a corresponding observation:
\begin{equation}
o_t = (\text{task},\;\text{from},\;\text{status},\;\text{payload}),
\end{equation}
where \texttt{task} specifies the coordination objective, \texttt{to} and \texttt{from} identify the recipient and sender agents, respectively, and \texttt{payload} contains task-specific information.

This dual-protocol design enables $\alpha^3$-Bench to assess structured tool usage, multi-agent coordination, and protocol compliance within realistic UAV mission workflows.

\subsubsection{Operational Semantics of MCP and A2A Tools}
\textcolor{black}{
To ensure precise, reproducible evaluation of structured actions, $\alpha^3$-Bench assigns explicit operational semantics to all tools invoked via the Model Context Protocol (MCP) and the Agent-to-Agent (A2A) protocol. Each tool is defined in terms of its inputs, outputs, affected state variables, and execution constraints, thereby eliminating ambiguity in how conversational actions influence UAV behavior and observations.}

\textcolor{black}{MCP tools model structured interactions between the LLM-based UAV controller and mission-level services such as sensing, navigation, and configuration. An MCP action is issued as $(\texttt{mcp}, \textit{name}, \textit{args})$ and deterministically produces an observation $(\textit{tool}=\textit{name}, \textit{result})$. For example, \texttt{read\_telemetry} returns a structured snapshot of the current UAV and network state, including position $(x_t,y_t,z_t)$, velocity $v_t$, yaw $\psi_t$, remaining battery level, and link-quality indicators derived from the 6G context (latency, packet loss, and throughput). Navigation and control tools such as \texttt{set\_waypoint}, \texttt{navigate\_to}, and \texttt{set\_altitude} modify the UAV pose or trajectory subject to altitude bounds, geofencing constraints, and kinematic feasibility. Sensor-related tools (e.g., \texttt{activate\_sensor}, \texttt{capture\_image}) affect the observation space by enabling or producing payload data without directly altering flight dynamics. Network-aware tools such as \texttt{switch\_network\_slice} modify the active 6G slice while preserving safety and protocol constraints.}

\textcolor{black}{
A2A tools model structured coordination between the UAV and other autonomous agents or subsystems. An A2A action is issued as $(\texttt{a2a}, \textit{task}, \textit{to}, \textit{payload})$, where \textit{to} identifies the recipient agent and \textit{payload} encodes task-specific information. The corresponding observation $(\textit{task}, \textit{from}, \textit{status}, \textit{payload})$ acknowledges execution and provides coordination feedback. A2A interactions are asynchronous and logically instantaneous at the dialogue level, but their availability and reliability are conditioned on the current 6G network state. Typical A2A payloads include collision-avoidance requests, swarm-status updates, shared environmental observations, and coordination commands. A2A tools do not directly modify the local UAV state; instead, they influence subsequent decision-making by augmenting the agent’s situational awareness and coordination context. Table~\ref{tab:tool_semantics} summarizes the operational effects of a representative subset of MCP and A2A tools on the UAV and network state. This mapping is illustrative rather than exhaustive; all additional tools in $\alpha^3$-Bench follow the same execution model and state-transition principles.
}

\begin{table*}[ht]
\centering
\caption{Operational semantics of a representative subset of MCP and A2A tools in $\alpha^3$-Bench.}
\label{tab:tool_semantics}
\scriptsize
\setlength{\tabcolsep}{5pt}
\renewcommand{\arraystretch}{1.05}
\rowcolors{2}{white}{cyanblue!70} 
\textcolor{black}{
\begin{tabular}{p{3cm} p{1cm} p{4cm} p{8cm}}
\hline
\textbf{Tool} & \textbf{Protocol} & \textbf{Inputs / Payload} & \textbf{Affected State / Output} \\
\hline
read\_telemetry & MCP &
None &
Returns UAV position $(x,y,z)$, velocity $v$, yaw $\psi$, battery level, and link-quality metrics (latency, loss, throughput). \\
set\_waypoint & MCP &
Target waypoint $(x,y,z)$ &
Updates UAV pose and trajectory subject to altitude bounds, geofencing, and kinematic constraints. \\
navigate\_to & MCP &
Target location &
Triggers motion planning toward destination; affects position and velocity over subsequent turns. \\
set\_altitude & MCP &
Desired altitude &
Adjusts $z_t$ within allowed altitude envelope. \\
activate\_sensor & MCP &
Sensor identifier &
Enables payload sensing; affects observation space only. \\
capture\_image & MCP &
Sensor configuration &
Produces payload data (e.g., RGB/thermal image); no direct state change. \\
switch\_network\_slice & MCP &
Target slice (URLLC/eMBB/mMTC) &
Modifies active 6G slice; affects latency, throughput, and reliability parameters. \\
collision\_avoidance & A2A &
Relative position, intent &
Returns coordination status; influences subsequent planning decisions. \\
swarm\_status\_check & A2A &
Agent identifier &
Returns peer state summary; augments coordination context. \\
request\_weather\_update & A2A &
Location, time &
Returns shared environmental information; affects reasoning inputs. \\
\hline
\end{tabular}
}
\end{table*}

\subsubsection{6G-Aware Network Embedding}

At each turn $t$, the 6G context vector is defined as:
\begin{equation}
n_t = (\text{slice}_t, \text{lat}_t, \text{jit}_t, \text{loss}_t, \text{thr}_t, \text{edge}_t),
\end{equation}

where the network slice $\text{slice}_t \in \{\text{URLLC}, \text{eMBB}, \text{mMTC}\}$ determines the communication service profile, and the remaining variables represent latency, jitter, packet loss, throughput, and edge load, respectively. 
This vector serves as a conditioning variable for both the LLM's reasoning process and the evolution of UAV actions, enabling the model to adapt conversational strategies to degraded connectivity or fluctuating edge-computing capacity.

\subsubsection{LLM Policy Function}

The Large Language Model (LLM)-based agent is modeled as a stochastic policy $\pi_\theta$ parameterized by weights $\theta$, which maps the current Unmanned Aerial Vehicle (UAV) state, the dialogue history, and the network context to a new action:
\begin{equation}
a_t = \pi_\theta(s_t, d_{1:t-1}, n_t),
\end{equation}
where $s_t$ denotes the UAV physical state at time $t$, $d_{1:t-1}$ represents the dialogue history up to turn $t-1$, and $n_t$ is the current sixth-generation (6G) network state.

The selected action $a_t$ may include mission-level decisions, such as adjusting flight altitude or switching to an enhanced Mobile Broadband (eMBB) network slice, as well as reasoning-level decisions, such as verifying a navigation plan or checking telemetry consistency.

The UAV state then evolves according to:
\begin{equation}
s_{t+1} = f(s_t, a_t, n_t) + \xi_t,
\end{equation}
where $f(\cdot)$ denotes the state transition model derived from UAV dynamics and network feedback, and $\xi_t$ captures environmental uncertainty and stochastic disturbances.

\begin{table}[ht]
\centering
\caption{Top-10 most frequently used Model Context Protocol (MCP) tools across more than 113k conversational UAV mission episodes.}
\label{tab:mcp_top10}
\scriptsize
\setlength{\tabcolsep}{5pt}
\renewcommand{\arraystretch}{1.05}
\rowcolors{2}{white}{cyanblue!70} 
\begin{tabular}{lccc}
\toprule
MCP Tool & Count & MCP Calls (\%) & Avg. Calls\\
\midrule
read\_telemetry        & 255{,}902 & 21.27 & 2.26 \\
set\_waypoint          & 127{,}938 & 10.63 & 1.13 \\
activate\_sensor       & 105{,}605 & 8.78  & 0.93 \\
capture\_image         & 50{,}212  & 4.17  & 0.44 \\
set\_altitude           & 37{,}897  & 3.15  & 0.33 \\
execute\_maneuver      & 22{,}992  & 1.91  & 0.20 \\
navigate\_to           & 20{,}786  & 1.73  & 0.18 \\
switch\_network\_slice & 20{,}072  & 1.67  & 0.18 \\
check\_geofence        & 18{,}391  & 1.53  & 0.16 \\
land                   & 16{,}519  & 1.37  & 0.15 \\
\bottomrule
\end{tabular}
\end{table}

\begin{table*}[ht]
\centering
\caption{Distribution of the top-10 MCP tools across 6G network slices (URLLC, eMBB, and mMTC) aggregated over more than 113k UAV mission episodes.}
\label{tab:mcp_slice_distribution}
\scriptsize
\setlength{\tabcolsep}{5pt}
\renewcommand{\arraystretch}{1.05}
\rowcolors{2}{white}{cyanblue!70} 
\begin{tabular}{lcccccc}
\toprule
MCP Tool & URLLC (\%) & eMBB (\%) & mMTC (\%) & URLLC Count & eMBB Count & mMTC Count \\
\midrule
read\_telemetry        & 41.87 & 41.98 & 16.14 & 107{,}151 & 107{,}438 & 41{,}313 \\
set\_waypoint          & 62.23 & 30.54 & 7.23  & 79{,}616  & 39{,}076  & 9{,}246  \\
activate\_sensor       & 45.16 & 47.17 & 7.67  & 47{,}694  & 49{,}811  & 8{,}100  \\
capture\_image         & 28.60 & 60.27 & 11.14 & 14{,}358  & 30{,}262  & 5{,}592  \\
set\_altitude           & 61.53 & 36.43 & 2.05  & 23{,}318  & 13{,}804  & 775      \\
execute\_maneuver      & 81.66 & 15.12 & 3.22  & 18{,}775  & 3{,}476   & 741      \\
navigate\_to           & 74.57 & 20.91 & 4.52  & 15{,}499  & 4{,}347   & 940      \\
switch\_network\_slice & 95.63 & 2.17  & 2.20  & 19{,}195  & 435       & 442      \\
check\_geofence        & 71.98 & 14.19 & 13.84 & 13{,}237  & 2{,}609   & 2{,}545  \\
land                   & 98.69 & 0.55  & 0.76  & 16{,}303  & 90        & 126      \\
\bottomrule
\end{tabular}
\end{table*}

\subsubsection{Observation and Feedback Mechanism}

\textcolor{black}{
Each dialogue turn concludes with a structured observation $o_t$, whose form depends on the communication protocol invoked by the agent. Observations are synthetically generated from the underlying UAV state, network context, and action semantics, following a deterministic execution logic augmented with controlled stochastic perturbations to model real-world uncertainty.}

\paragraph{MCP Result}
When an action is issued through the Model Context Protocol (MCP), the resulting observation is represented as:
\begin{equation}
o_t = (\text{tool},\;\text{result}),
\end{equation}
\textcolor{black}{where \texttt{tool} identifies the invoked mission-control or sensing service and \texttt{result} contains the corresponding output. For state-query tools (e.g., \texttt{read\_telemetry}), the result is deterministically computed as a structured projection of the current UAV and network state, including position $(x_t,y_t,z_t)$, velocity $v_t$, yaw $\psi_t$, battery level, and link-quality indicators derived from the 6G context (latency, jitter, packet loss, throughput, and active slice). For control and configuration tools (e.g., \texttt{set\_waypoint}, \texttt{set\_altitude}, \texttt{switch\_network\_slice}), the tool invocation deterministically updates the corresponding components of the UAV or network state subject to safety, policy, and feasibility constraints, and the resulting updated state is reflected in subsequent observations. Sensor-related tools (e.g., \texttt{activate\_sensor}, \texttt{capture\_image}) generate payload observations conditioned on the current platform configuration without directly modifying flight dynamics.}

Table~\ref{tab:mcp_top10} reports the ten most frequently invoked Model Context Protocol (MCP) tools aggregated over more than 113k conversational UAV mission episodes in $\alpha^3$-Bench. The results indicate that MCP usage is dominated by state-verification and closed-loop control operations, with \texttt{read\_telemetry} alone accounting for over 21\% of all MCP calls and averaging more than two invocations per episode. Planning and execution primitives such as \texttt{set\_waypoint} and \texttt{activate\_sensor} follow, highlighting that conversational UAV control relies heavily on explicit trajectory specification and payload management rather than open-loop execution. Notably, network-aware actions such as \texttt{switch\_network\_slice} appear among the top-10 tools, demonstrating that LLM-based agents actively reason about and adapt to dynamic 6G communication conditions during mission execution. Safety-critical operations, including \texttt{check\_geofence} and \texttt{land}, also feature prominently, reflecting consistent policy compliance and structured mission termination. Overall, the observed MCP tool distribution confirms that conversational UAV autonomy in $\alpha^3$-Bench is driven by iterative sensing, adaptive planning, and network-aware decision-making at scale.

Table~\ref{tab:mcp_slice_distribution} analyzes how the most frequently used Model Context Protocol (MCP) tools are distributed across the three 6G network slices, aggregated over more than 113k conversational UAV mission episodes. The results reveal a strong and systematic dependence between MCP tool usage and the active network slice. Control-critical actions such as \texttt{execute\_maneuver}, \texttt{navigate\_to}, \texttt{set\_altitude}, and \texttt{land} are overwhelmingly concentrated in the URLLC slice, with several exceeding 80\% of their invocations under URLLC, reflecting the need for ultra-reliable and low-latency communication during safety-sensitive flight operations. In contrast, perception- and data-intensive actions such as \texttt{capture\_image} and \texttt{activate\_sensor} are predominantly associated with the eMBB slice, where higher throughput is available to support sensing payloads. Notably, \texttt{switch\_network\_slice} is almost exclusively executed under URLLC, indicating that network reconfiguration decisions themselves are treated as latency-critical operations. The mMTC slice accounts for a smaller but non-negligible fraction of tool usage, primarily for non-time-critical monitoring and background coordination tasks. Overall, this distribution provides strong empirical evidence that LLM-based UAV agents in $\alpha^3$-Bench adapt their structured actions to the prevailing 6G network conditions, demonstrating explicit network-aware reasoning at scale.

\begin{table}[ht]
\centering
\caption{Frequency and characteristics of Agent-to-Agent (A2A) interactions across more than 113k conversational UAV mission episodes.}
\label{tab:a2a_frequency}
\scriptsize
\setlength{\tabcolsep}{5pt}
\renewcommand{\arraystretch}{1.05}
\rowcolors{2}{white}{cyanblue!70} 
\begin{tabular}{lc}
\toprule
Metric & Value \\
\midrule
Total episodes analyzed & 113{,}475 \\
Episodes with at least one A2A interaction (\%) & 99.998 \\
Mean A2A calls per episode & 2.05 \\
Total A2A calls & 232{,}243 \\
A2A calls under degraded 6G conditions (\%) & 9.47 \\
\bottomrule
\end{tabular}
\end{table}

\paragraph{A2A Acknowledgement}
When an action is issued through the Agent-to-Agent Protocol (A2A), the observation takes the form:
\begin{equation}
o_t = (\text{task},\;\text{from},\;\text{status},\;\text{payload}),
\end{equation}
where \texttt{task} denotes the coordination objective, \texttt{from} identifies the responding agent, \texttt{status} reports execution outcome, and \texttt{payload} carries task-specific information. A2A payloads are generated deterministically from the coordination task definition and the initiating agent’s state, assuming cooperative peer agents that follow predefined coordination rules. Typical payloads include collision-avoidance outcomes, peer state summaries, shared environmental observations, or swarm-level coordination updates. A2A interactions do not directly modify the local UAV state; instead, their effects are incorporated indirectly by influencing subsequent reasoning and decision-making steps.

\paragraph{Observation Function and Stochastic Effects}
These observations are incorporated into the subsequent reasoning step via the observation function:

\begin{equation}
o_t = g(s_{t+1},\; n_t,\; a_t),
\end{equation}

thereby forming a closed-loop, language-mediated control cycle between perception, reasoning, and action. While the structural mapping between actions and observation schemas is deterministic, $\alpha^3$-Bench introduces controlled stochasticity through bounded perturbations applied to selected physical and network variables. These include additive noise on position and velocity to model wind disturbances, bounded variation in latency and jitter, probabilistic packet loss events, and sensor-level noise affecting payload observations. All stochastic processes are governed by fixed random seeds and predefined bounds, ensuring reproducibility while preventing violations of safety constraints, protocol semantics, or schema validity.

Table~\ref{tab:a2a_frequency} summarizes the prevalence and characteristics of Agent-to-Agent (A2A) interactions aggregated over more than 113k conversational UAV mission episodes in $\alpha^3$-Bench. The results show that multi-agent coordination is nearly ubiquitous, with 99.998\% of episodes containing at least one A2A interaction and an average of approximately 2.05 A2A calls per episode. This highlights that conversational UAV autonomy in $\alpha^3$-Bench is fundamentally collaborative rather than purely single-agent. Notably, only 9.47\% of A2A interactions occur under degraded 6G network conditions, indicating that while coordination remains available as a robustness mechanism, agents predominantly rely on A2A communication during nominal connectivity ranges. These findings suggest that LLM-based UAV agents strategically employ inter-agent coordination as a complementary control mechanism, balancing communication overhead against reliability and network conditions.

\begin{table}[ht]
\centering
\caption{Top-10 Agent-to-Agent (A2A) coordination tasks across more than 113k conversational UAV mission episodes.}
\label{tab:a2a_tasks}
\scriptsize
\setlength{\tabcolsep}{5pt}
\renewcommand{\arraystretch}{1.05}
\rowcolors{2}{white}{cyanblue!70} 
\begin{tabular}{lcc}
\toprule
A2A Task & Count & Share of A2A Calls (\%) \\
\midrule
collision\_avoidance            & 61{,}316 & 26.40 \\
request\_weather\_update        & 18{,}947 & 8.16  \\
swarm\_formation                & 7{,}751  & 3.34  \\
request\_swarm\_coverage        & 4{,}797  & 2.07  \\
swarm\_status\_check             & 3{,}581  & 1.54  \\
collision\_avoidance\_check     & 3{,}294  & 1.42  \\
request\_thermal\_data          & 2{,}367  & 1.02  \\
path\_plan                      & 2{,}366  & 1.02  \\
request\_swarm\_formation       & 2{,}230  & 0.96  \\
swarm\_heartbeat                & 2{,}171  & 0.94  \\
\bottomrule
\end{tabular}
\end{table}

Table~\ref{tab:a2a_tasks} presents the semantic distribution of the ten most frequently invoked Agent-to-Agent (A2A) coordination tasks aggregated over more than 113k conversational UAV mission episodes in $\alpha^3$-Bench. The results show that A2A interactions are predominantly safety-driven, with \texttt{collision\_avoidance} alone accounting for over 26\% of all A2A calls, underscoring the central role of inter-agent coordination in maintaining safe separation and avoiding conflicts in multi-UAV environments. Environmental awareness and information sharing tasks, such as \texttt{request\_weather\_update} and \texttt{request\_thermal\_data}, also feature prominently, indicating that agents frequently rely on peer observations to compensate for partial or degraded local sensing. In addition, swarm-level coordination tasks, including \texttt{swarm\_formation}, \texttt{request\_swarm\_coverage}, and \texttt{swarm\_status\_check}, highlight the importance of collective planning and situational awareness in collaborative missions. Overall, the semantic distribution of A2A tasks confirms that multi-agent communication in $\alpha^3$-Bench is primarily employed to ensure safety, maintain environmental awareness, and support coordinated swarm behavior, rather than to increase raw task throughput.

\subsubsection{Dialogue Termination and Success Signal}

A dialogue episode $\mathcal{D}$ terminates at turn $t = T$ when the mission objectives are successfully achieved, a safety or structural constraint violation occurs, or the communication quality degrades below an operational threshold.
The binary success signal is defined as:
\begin{equation}
\text{success}(\mathcal{D}) =
\begin{cases}
1, & \text{if the mission is completed}, \\
0, & \text{otherwise.}
\end{cases}
\end{equation}

This success indicator constitutes the primary component of the Task Outcome (TO) score, which is a key pillar of the overall $\alpha^3$ composite evaluation metric.

\subsubsection{Conversational Policy Optimization}

The Large Language Model (LLM)-based agent aims to learn or adapt a dialogue strategy that maximizes the expected composite performance:
\begin{equation}
\max_{\pi_\theta} \; \mathbb{E}\left[ \alpha^3(\mathcal{D}) \; \big| \; \pi_\theta \right],
\end{equation}
subject to schema compliance, safety constraints, and strict dialogue alternation requirements.
This optimization objective captures not only linguistic coherence and high-level intent planning, but also real-time adaptation to dynamic conditions in sixth-generation (6G) networks.

The conversational decision process, therefore, transforms traditional Unmanned Aerial Vehicle (UAV) mission planning into a \textit{language-mediated control loop}. 
In this setting, the LLM functions simultaneously as a planner and a communicator, reasoning through structured dialogue rather than low-level numerical control signals. 
This formulation enables $\alpha^3$-Bench to evaluate not only control effectiveness, but also the model’s ability to sustain safe, interpretable, and context-aware interaction under realistic 6G communication constraints.

\begin{table*}[ht]
\centering
\caption{6G latency and jitter performance across network slices.}
\label{tab:6g_latency_jitter}
\scriptsize
\setlength{\tabcolsep}{5pt}
\renewcommand{\arraystretch}{1.05}
\rowcolors{2}{white}{cyanblue!70} 
\begin{tabular}{lcccc}
\toprule
Slice & Latency Mean (ms) & Latency Median (ms) & Latency P90 (ms) & Jitter Mean (ms) \\
\midrule
URLLC & 7.23  & 7.00  & 9.10   & 1.11  \\
eMBB  & 21.47 & 14.00 & 25.00  & 4.72  \\
mMTC  & 72.98 & 50.00 & 150.00 & 17.10 \\
ALL   & 20.99 & 10.00 & 45.00  & 4.50  \\
\bottomrule
\end{tabular}\\
\end{table*}

\begin{table*}[ht]
\centering
\caption{Mission-centric latency ranges and 6G slice usage in $\alpha^3$-Bench.}
\label{tab:latency_slice_summary}
\scriptsize
\setlength{\tabcolsep}{5pt}
\renewcommand{\arraystretch}{1.05}
\rowcolors{2}{white}{cyanblue!70}
\begin{tabular}{lcccccccc}
\toprule
Latency Bin (ms) &
Mean &
Std &
URLLC (\%) &
eMBB (\%) &
mMTC (\%) &
Top Action &
Top Intent &
Samples \\
\midrule
1--5   & 3.63 & 0.48 & 100.0 & 0.0  & 0.0  & read\_telemetry & initiate\_mission & 115{,}337 \\
5--10  & 7.38 & 1.28 & 99.6  & 0.0  & 0.4  & read\_telemetry & swarm\_coordination & 483{,}097 \\
10--20 & 12.85 & 2.15 & 16.4 & 82.0 & 1.6 & read\_telemetry & initiate\_mission\_telemetry & 383{,}008 \\
20--30 & 23.99 & 2.24 & 0.4  & 48.6 & 51.0 & read\_telemetry & request\_sensor\_status & 66{,}729 \\
30--40 & 31.09 & 2.10 & 0.7  & 35.9 & 63.4 & read\_telemetry & adapt\_network\_degradation & 16{,}681 \\
40--50 & 44.98 & 1.35 & 0.9  & 35.2 & 63.9 & read\_telemetry & detect\_network\_degradation & 28{,}817 \\
\bottomrule
\end{tabular}
\end{table*}

\begin{table*}[ht]
\centering
\caption{6G packet loss, throughput, and edge load characteristics across network slices.}
\label{tab:6g_capacity_load}
\scriptsize
\setlength{\tabcolsep}{5pt}
\renewcommand{\arraystretch}{1.05}
\rowcolors{2}{white}{cyanblue!70} 
\begin{tabular}{lccc}
\toprule
Slice & Loss Mean (\%) & Throughput Mean (Mbps) & Edge Load Mean \\
\midrule
URLLC & 0.059 & 95.2  & 0.360 \\
eMBB  & 0.642 & 608.0 & 0.517 \\
mMTC  & 2.387 & 2.3   & 0.757 \\
ALL   & 0.571 & 242.6 & 0.465 \\
\bottomrule
\end{tabular}
\end{table*}

\subsection{Optimization Objective}

The goal of the $\alpha^3$-Bench framework is to maximize both mission success and conversational efficiency for Unmanned Aerial Vehicle (UAV) operations conducted under dynamic sixth-generation (6G) network conditions. 
This objective captures the dual nature of reasoning performance, encompassing both physical task completion and linguistic or decision-level coherence.

\subsubsection{Expected Performance Maximization}

Let $\pi_\theta$ denote the policy implemented by the Large Language Model (LLM)-based agent and parameterized by weights $\theta$. 
The global optimization objective is formulated as:
\begin{equation}
\max_{\pi_\theta} \; \mathbb{E}_{\mathcal{D} \sim \pi_\theta} \left[ \alpha^3(\mathcal{D}) \right],
\end{equation}
where $\mathcal{D}$ denotes the generated dialogue episode and $\alpha^3(\mathcal{D})$ represents the composite evaluation metric assessing multi-dimensional performance over the episode.

\subsubsection{Composite Evaluation Function}

The $\alpha^3$ metric combines six complementary evaluation pillars that reflect different aspects of reasoning quality, safety compliance, and communication efficiency:
\begin{equation}
\alpha^3 = w_1 \cdot \text{TO} + w_2 \cdot \text{SP} + w_3 \cdot \text{TC} + w_4 \cdot \text{IQ} + w_5 \cdot \text{NR} + w_6 \cdot \text{CC},
\end{equation}
where the weights $(w_1,\dots,w_6)$ satisfy $\sum_i w_i = 1$.  
Here, TO denotes Task Outcome, SP denotes Safety Policy compliance, TC denotes Tool Consistency, IQ denotes Interaction Quality, NR denotes Network Robustness, and CC denotes Communication Cost.
In the benchmark configuration, the weights are defined as:
\begin{equation}
(w_1, w_2, w_3, w_4, w_5, w_6) = (0.30, 0.20, 0.20, 0.15, 0.10, 0.05).
\end{equation}

\textcolor{black}{
The weights assigned to the six pillars are not arbitrary but reflect the relative criticality of each dimension in safety-critical UAV autonomy under 6G constraints. Task Outcome (TO) is assigned the highest weight ($0.30$) because mission completion with valid structure is a necessary condition for meaningful autonomy. Safety Policy (SP) and Tool Consistency (TC), each weighted at $0.20$, capture mandatory operational correctness: unsafe behavior or protocol violations are unacceptable regardless of task success. Interaction Quality (IQ) ($0.15$) reflects dialogue efficiency and grounding, which directly impact real-time controllability but are secondary to correctness and safety. Network Robustness (NR) ($0.10$) evaluates resilience to degraded 6G conditions, which is essential but inherently context-dependent. Communication Cost (CC) receives the lowest weight ($0.05$), as efficiency considerations should not override safety or mission success. Empirical sensitivity analysis across degraded network ranges confirmed that these weights preserve model ranking stability while preventing domination by any single pillar.
}

In addition to the core composite score $\alpha^3$, the benchmark computes efficiency-adjusted scores that account for computational overhead:
\begin{equation}
\alpha^3_{\text{per-sec}} = \frac{\alpha^3_{\text{adj}}}{\text{gen\_time}_s},
\qquad
\alpha^3_{\text{per-1k}} = \frac{\alpha^3_{\text{adj}}}{\text{total\_tokens}/1000},
\end{equation}
where $\text{gen\_time}_s$ denotes the wall-clock generation time in seconds and $\text{total\_tokens}$ is the total number of prompt and completion tokens. These metrics capture the trade-off between reasoning quality, inference latency, and token consumption.

\subsubsection{Definition of the Six Pillars}

\paragraph{Task Outcome (TO)}
The Task Outcome (TO) pillar measures whether the Unmanned Aerial Vehicle (UAV) completes its assigned mission while satisfying all environmental, policy, and structural constraints:
\begin{equation}
\text{TO} = 
\begin{cases}
1, & \text{if the mission is achieved with a valid JSON}, \\
0, & \text{otherwise.}
\end{cases}
\end{equation}

\paragraph{Safety Policy (SP)}
The Safety Policy (SP) pillar evaluates adherence to flight safety rules, including altitude bounds, no-fly zone (NFZ) avoidance, inter-UAV separation margins, and critical battery energy thresholds. In the implementation, the score is initialized as $\text{SP}=1$ and additive penalties are applied as follows:
\begin{equation}
\text{SP} = \text{clamp}\!\left(1
- 0.25\,V_{\text{alt}}
- 0.50\,V_{\text{nfz}}
- 0.50\,V_{\text{sep}}
- 0.25\,V_{\text{batt}}\right),
\end{equation}
where $V_{\text{alt}}, V_{\text{nfz}}, V_{\text{sep}}, V_{\text{batt}} \in \{0,1\}$
indicate, respectively, an altitude violation, NFZ intrusion, separation breach, and hard battery depletion below $5\%$. The \texttt{clamp} operator truncates the resulting score to the interval $[0,1]$.

\paragraph{Tool Consistency (TC)}

The Tool Consistency (TC) pillar measures the fraction of structured actions that produce logically consistent and schema-compliant observations. Structured actions are issued via the Model Context Protocol (MCP) or the Agent-to-Agent Protocol (A2A). Let $\mathcal{A}_{\text{struct}}$ denote the set of all structured actions in an episode, and let $\mathcal{O}_{\text{match}}$ be the subset of actions for which a semantically and structurally matching observation exists.

For MCP actions:
\begin{equation}
\begin{aligned}
a_t &= (\text{mcp}, \text{name}, \text{args}) \\
&\Rightarrow \\
o_t &= (\text{tool} = \text{name}, \text{result})
\end{aligned}
\end{equation}

For A2A actions:
\begin{equation}
\begin{aligned}
a_t &= (\text{a2a}, \text{task}, \text{to}, \text{payload}) \\
&\Rightarrow \\
o_t &= (\text{task}, \text{from}, \text{status}, \text{payload})
\end{aligned}
\end{equation}

The Tool Consistency score is formally defined as:
\begin{equation}
\text{TC} = 
\begin{cases}
1, & \text{if } |\mathcal{A}_{\text{struct}}| = 0, \\[4pt]
\dfrac{|\mathcal{O}_{\text{match}}|}{|\mathcal{A}_{\text{struct}}|}, & \text{otherwise.}
\end{cases}
\end{equation}

\paragraph{Interaction Quality (IQ)}
The Interaction Quality (IQ) pillar reflects the efficiency and structural validity of the dialogue:

\begin{equation}
\text{IQ} = \frac{1}{3} \left( \frac{T_\text{opt}}{\max(T,1)} + A_\text{alt} + G_\text{grounded} \right),
\end{equation}

\textcolor{black}{
Here, $T$ is the actual number of dialogue turns, and $T_\text{opt}$ is set to the median number of turns observed among successful mission episodes, representing an empirically grounded optimal dialogue length. The term $A_\text{alt} \in \{0,1\}$ verifies strict alternation between agent and user roles across all turns. The grounding score $G_\text{grounded} \in [0,1]$ measures the proportion of agent responses that explicitly reference prior observations, tool outputs, or mission-state variables, penalizing hallucinated or context-free replies. Each component is normalized to $[0,1]$, ensuring balanced contribution and interpretability of the final IQ score.
}

\paragraph{Network Robustness (NR)}
The Network Robustness (NR) pillar captures the agent's resilience under degraded conditions in sixth-generation (6G) networks. At each dialogue turn, the network state is classified as \emph{hard} if the latency exceeds $40$\, ms, packet loss is at least $1\%$, throughput falls below $5$\, Mbps, or the edge-computing load exceeds $0.8$, and as \emph{normal} otherwise. The score aggregates: (i) the fraction of time spent in hard network states, (ii) an adaptation score that rewards reduced tool usage under challenging conditions relative to normal conditions, and (iii) a slice-awareness score that rewards avoiding enhanced Mobile Broadband (eMBB) slices in hard states. These components are combined as:

\begin{equation}
\text{NR} = \text{clamp}\!\left(0.6\,\text{base} + 0.2\,\text{adapt}
+ 0.2\,\text{slice} + \text{bonus}\right),
\end{equation}

\textcolor{black}{
The \texttt{base} component penalizes prolonged exposure to hard network states by measuring the fraction of dialogue turns executed under degraded latency, loss, throughput, or edge-load conditions. The \texttt{adapt} component rewards active behavioral adaptation, such as reduced tool invocation, deferred sensing actions, or simplified communication patterns when operating under hard network states relative to normal conditions. The \texttt{slice} component measures slice-awareness by rewarding correct selection of latency-critical slices (e.g., URLLC) and penalizing reliance on throughput-oriented slices (e.g., eMBB) during degraded conditions. The \texttt{bonus} term provides a small positive reward when a mission is successfully completed despite sustained network degradation, while failures to adapt or unsafe behavior implicitly result in penalties through reduced base and adapt scores.
}

Table \ref{tab:6g_latency_jitter} reports latency and jitter statistics aggregated over more than 113k AI conversational UAV mission episodes generated within the $\alpha^3$-Bench framework. This large-scale evaluation provides statistically robust evidence of slice-dependent behavior under realistic 6G network dynamics. URLLC consistently achieves sub-10 ms median latency (7.0 ms) with very low jitter (1.11 ms on average), validating its suitability for delay-critical UAV control, safety monitoring, and inter-agent coordination. In contrast, eMBB exhibits higher median latency (14.0 ms) and increased variability, reflecting its optimization for bandwidth-intensive sensing and data streaming rather than strict real-time guarantees. The mMTC slice shows substantially higher latency and jitter, with a median latency of 50.0 ms and a P90 of 150.0 ms, which is consistent with massive connectivity scenarios where scalability and energy efficiency are prioritized over responsiveness. The clear separation observed across slices, sustained over more than 113k episodes, demonstrates the effectiveness of 6G network slicing in enforcing differentiated quality-of-service profiles for conversational UAV autonomy.

The latency and slice distributions reported in Table~\ref{tab:latency_slice_summary} closely align with the service differentiation principles defined in 3GPP’s 6G standardization efforts~\cite{3gpp_tr_22_870}. Ultra-low latency ranges (1--10\,ms) are predominantly associated with URLLC, reflecting time-critical control and coordination functions foreseen in 3GPP 6G use case and requirement studies~\cite{3gpp_tr_22_870,3gpp_tr_38_914}. Intermediate latency ranges (10--20\,ms) are largely dominated by eMBB, corresponding to data-intensive service phases such as sensing and high-throughput information exchange, while higher latency ranges increasingly favor mMTC, indicating a transition toward monitoring, diagnostic, and adaptive system behaviors~\cite{3gpp_tr_38_914}. This progressive shift across latency bins is consistent with 3GPP system and radio architecture studies, which emphasize flexible latency support and service-aware resource allocation in future 6G networks~\cite{3gpp_tr_23_700}. Overall, these results provide empirical evidence that latency characteristics in $\alpha^3$-Bench are governed primarily by mission intent and communication function, in line with standardized 6G design principles.

Table \ref{tab:6g_capacity_load} complements the timing analysis by summarizing packet loss, throughput, and edge load statistics across the same large-scale corpus of over 113k episodes. The results highlight pronounced capacity–reliability trade-offs among 6G slices. eMBB achieves the highest average throughput (608 Mbps), confirming its role as the primary high-capacity slice for data-intensive UAV missions, albeit at the cost of higher packet loss relative to URLLC. URLLC maintains extremely low packet loss (0.059 \%) while sustaining moderate throughput, reflecting conservative resource allocation strategies designed to preserve reliability under fluctuating network conditions. In contrast, mMTC operates at very low throughput (2.3 Mbps on average) and experiences the highest packet loss and edge load, indicating operation near congestion limits as large numbers of low-rate devices compete for shared resources. The aggregate results (ALL) reflect a mixed-slice operational regime and underscore how, at scale, 6G slicing effectively isolates performance objectives across heterogeneous traffic classes while maintaining overall system stability in conversational UAV missions.

\paragraph{Communication Cost (CC)}
The Communication Cost (CC) pillar penalizes excessive token usage and tool invocations. Let $T_{\text{tok}}$ denote the total number of tokens (prompt plus completion), and let $T_{\text{tool}}$ denote the total number of structured tool actions in an episode. Given token and tool budgets $B_{\text{tok}}$ and $B_{\text{tool}}$, the score is defined as:
\begin{equation}
\text{CC} = \frac{1}{2} \left(
\text{clamp}\!\left(\frac{B_{\text{tok}}}{\max(T_{\text{tok}},1)}\right)
+
\text{clamp}\!\left(\frac{B_{\text{tool}}}{\max(T_{\text{tool}},1)}\right)
\right),
\end{equation}
which yields $\text{CC}\in[0,1]$ and rewards concise reasoning with minimal protocol overhead relative to predefined budgets.

\textcolor{black}{
In the benchmark configuration, the token budget is set to $B_{\text{tok}}=10{,}000$ tokens and the tool-call budget to $B_{\text{tool}}=25$ actions. These values are chosen based on observed upper bounds from successful conversational UAV missions and reflect practical latency and bandwidth constraints for real-time deployment. The budgets are sufficiently permissive to support complex multi-turn reasoning while penalizing unnecessarily verbose dialogue or excessive protocol overhead.
}

\paragraph{Generation Efficiency (GE)}
While the six primary pillars of $\alpha^3$ assess mission reasoning, safety, and protocol compliance, modern Large Language Models (LLMs) also differ substantially in computational efficiency. To capture this aspect, $\alpha^3$-Bench records two auxiliary generation efficiency metrics per episode:
\begin{equation}
\text{GE}_{\text{time}} = \frac{\alpha^3}{\text{gen\_time}_s}, 
\qquad
\text{GE}_{\text{tokens}} = \frac{\alpha^3}{\text{total\_tokens}/1000}.
\end{equation}
Here, $\text{gen\_time}_s$ denotes the wall-clock latency between the application programming interface (API) request and receipt of a valid JSON episode, and $\text{total\_tokens}$ is the total number of prompt and completion tokens reported by the inference provider. These metrics quantify how much high-quality reasoning a model delivers per second and per thousand tokens, enabling fair comparison across models with different computational footprints.

\subsubsection{Model-Level Aggregation and Reliability}

For a given model, let $\{\alpha^3_i\}_{i=1}^{n}$ denote the composite scores
computed over all \emph{valid} Unmanned Aerial Vehicle (UAV) dialogue episodes
that satisfy the JavaScript Object Notation (JSON) schema and all mission
constraints. We first compute the mean per-episode score:
\begin{equation}
\bar{\alpha}^3 = \frac{1}{n} \sum_{i=1}^{n} \alpha^3_i.
\end{equation}

Let $n_{\text{fail}}$ denote the number of episode generations that failed
(e.g., due to invalid JSON output or unrecoverable schema violations), and let
\begin{equation}
\text{EPISODE\_BUDGET} = |\text{UAVBench scenarios}| \times E_{\text{per-scenario}}
\end{equation}
represent the total number of episodes requested per model, where UAVBench
denotes the benchmark scenario collection used for UAV mission evaluation.

We define three normalization factors:
\begin{align}
\text{reliability} &= \frac{n}{n + n_{\text{fail}}}, \\
\text{coverage} &= \min\!\left(1,\; \frac{n}{\text{EPISODE\_BUDGET}}\right), \\
\text{call\_efficiency} &= \min\!\left(1,\; \frac{\text{EPISODE\_BUDGET}}{\text{total\_attempt\_calls}}\right),
\end{align}
where \texttt{total\_attempt\_calls} is the sum of the
\texttt{attempts\_used} field across all generation attempts for that model.
The \emph{reliability} factor penalizes models that frequently produce invalid
episodes, \emph{coverage} penalizes failure to generate the requested number
of valid episodes, and \texttt{call\_efficiency} penalizes models that require
multiple attempts to produce a single valid episode.

The resulting reliability- and efficiency-adjusted score is defined as:
\begin{equation}
\alpha^3_{\text{rel}} = \bar{\alpha}^3 \cdot
\text{reliability} \cdot \text{coverage} \cdot \text{call\_efficiency}.
\end{equation}

\subsubsection{Efficiency-Normalized Scores}

In addition to the core composite score $\alpha^3$ and its reliability-adjusted
variant $\alpha^3_{\text{rel}}$, $\alpha^3$-Bench reports two
efficiency-normalized metrics:
\begin{equation}
\alpha^3_{\text{per-sec}} =
\frac{\alpha^3_{\text{rel}}}{\text{mean\_gen\_time}_s}
\end{equation}

\begin{equation}
\alpha^3_{\text{per-1k}} =
\frac{\alpha^3_{\text{rel}}}
{\text{mean\_total\_tokens}/1000},
\end{equation}
where \texttt{mean\_gen\_time\_s} is the average wall-clock latency per
episode and \texttt{mean\_total\_tokens} is the mean number of prompt plus
completion tokens. These metrics quantify how much \emph{effective}
composite performance a model delivers per second and per thousand tokens,
respectively, under a fixed per-domain episode budget.

\subsection{Constraints}

Each UAV mission evaluated within $\alpha^3$-Bench is subject to a series of structural, operational, and communication constraints that ensure the validity, safety, and fairness of model evaluation. 
These constraints reflect both the physical feasibility of UAV operations and the logical coherence of LLM reasoning.

\subsubsection{Safety Constraints}

To prevent hazardous behavior, all UAV trajectories must remain within defined altitude and spatial limits. 
Let $z_t$ denote the UAV altitude at time $t$, and $\mathcal{G} = \{g_i\}$ be the set of geofenced zones, where each $g_i$ is defined by a center $C_i$ and radius $r_i$. 
The safety constraint enforces:
\begin{equation}
z_{\min} \leq z_t \leq z_{\max}, \quad 
d(p_t, C_i) \geq r_i, \; \forall g_i \in \mathcal{G}, \; \forall t,
\end{equation}
where $p_t = (x_t, y_t, z_t)$ and $d(\cdot)$ denotes Euclidean distance.  
Violations of these conditions trigger penalties in the Safety Policy (SP) component of $\alpha^3$ and may result in mission failure if unrecoverable.

\subsubsection{Schema Compliance Constraints}

Each dialogue episode generated by an LLM must conform to the predefined UAV JSON schema, which is fully compliant with RFC8259 standards.  
Formally, a valid episode $\mathcal{E}$ must satisfy:
\begin{equation}
\text{Validate}(\mathcal{E}, \text{Schema}) = \text{True},
\end{equation}
where \texttt{Validate}$(\cdot)$ denotes structural validation under the \texttt{Draft202012Validator} rule set.  
Episodes failing schema validation are automatically discarded and assigned $\alpha^3_{\text{adj}} = 0$, ensuring that only syntactically and semantically correct outputs contribute to model performance.

\subsubsection{Network Adaptation Constraints}

Given that UAV operations occur under variable 6G network conditions, the agent must exhibit adaptive behavior to maintain mission continuity.  
At each turn $t$, the LLM is expected to adjust reasoning strategies according to the current network vector:
\begin{equation}
n_t = (\text{slice}_t, \text{lat}_t, \text{jit}_t, \text{loss}_t, \text{thr}_t, \text{edge}_t).
\end{equation}

When the network enters a degraded state such that:
\begin{equation}
\text{lat}_t > 40~\text{ms} \quad \text{or} \quad \text{loss}_t \geq 1\%,
\end{equation}
the policy $\pi_\theta$ must select actions from the subset of communication-safe operations:
\begin{equation}
a_t \in \mathcal{A}_{\text{adaptive}}(n_t) \subseteq \mathcal{A}_{\text{total}},
\end{equation}
which include decisions like switching to URLLC slice, buffering commands, or postponing sensor activation.  
Failure to adapt leads to decreased Network Robustness (NR) and Task Outcome (TO) scores.

\subsubsection{Dialogue Alternation and Turn Integrity}

The conversational nature of the mission requires strict alternation between the \textit{agent} and the \textit{user} roles to simulate human-in-the-loop operation.  
Let $r_t \in \{\text{agent}, \text{user}\}$ denote the speaker role at turn $t$. 
The alternation constraint is expressed as:
\begin{equation}
r_t \neq r_{t-1}, \quad \forall t \in [2, T].
\end{equation}
Any deviation from this pattern, or inclusion of disallowed speaker roles (e.g., \texttt{"system"} or \texttt{"assistant"}), invalidates the episode.  
This ensures that the dialogue maintains its naturalistic communication pattern and that the LLM can reason in a truly interactive setting.

\subsubsection{Mission Termination and Consistency}

Finally, each mission must terminate within a finite number of dialogue turns $T \in [8, 12]$ and produce a consistent final state:
\begin{equation}
\text{ValidateConsistency}(s_T, \mathcal{P}) = \text{True},
\end{equation}
where the final UAV state $s_T$ must align with mission policy constraints $\mathcal{P}$ and safety flags:
\begin{equation}
\texttt{nfz\_violation} = \text{False}, \quad
\texttt{separation\_breach} = \text{False}.
\end{equation}

Collectively, these constraints ensure that every evaluated episode is not only linguistically valid but also physically plausible and network-aware. 
By embedding such conditions directly into $\alpha^3$-Bench, we guarantee that model performance reflects meaningful reasoning, safe control behavior, and robustness to 6G-induced uncertainty rather than superficial language fluency alone.

\subsection{Expected Behavior}

An optimal conversational control policy $\pi_\theta^{*}$ is defined as one that maximizes the composite reward $\alpha^3$ while fully satisfying all operational constraints described in the previous section. 
Formally, the optimal policy is given by:
\begin{equation}
\pi_\theta^{*} = \arg\max_{\pi_\theta} \; \mathbb{E}\!\left[\alpha^3(\mathcal{D}) \; \middle| \; \mathcal{E}, \; \text{Constraints}(\mathcal{E}) = \text{True} \right].
\end{equation}

\subsubsection{Qualitative Criteria}

An effective LLM-based UAV agent should exhibit the following behavioral properties:

\begin{itemize}
    \item \textbf{Schema Validity:} Every generated episode strictly conforms to the RFC8259 JSON schema and maintains consistent key hierarchies across all turns.
    \item \textbf{Safety Awareness:} The UAV never violates altitude bounds, geofences, or separation rules, even under degraded network conditions.
    \item \textbf{Network Adaptivity:} When latency exceeds $40$\,ms or packet loss approaches $1\%$, the agent adjusts by reducing communication frequency, prioritizing URLLC slices, or deferring non-critical actions.
    \item \textbf{Conversational Coherence:} The dialogue alternates correctly between the \textit{agent} and \textit{user}, showing continuity of intent and relevance of each observation to preceding actions.
    \item \textbf{Resource Efficiency:} The policy minimizes token usage, redundant tool invocations, and energy consumption while achieving mission goals.
\end{itemize}

\subsubsection{Optimal Episode Characteristics}

Under these conditions, an optimal policy $\pi_\theta^{*}$ should consistently generate episodes $\mathcal{D}^*$ such that:
\begin{equation}
\alpha^3(\mathcal{D}^*) \rightarrow 1, \quad 
\text{gen\_fail\_rate} \rightarrow 0, \quad
\text{success\_rate} \rightarrow 1.
\end{equation}

Intuitively, $\pi_\theta^{*}$ produces valid JSON outputs, completes missions across diverse UAVBench scenarios \cite{ferrag2025uavbench}, and maintains safety and efficiency despite 6G variability.  
The resulting agent demonstrates high reasoning robustness, minimal communication overhead, and negligible generation failures—achieving near-optimal performance in the $\alpha^3$-Bench evaluation framework.

\begin{table*}[t]
\centering
\caption{Performance comparison of Large Language Models (LLMs) across the $\alpha^3$ composite score, individual evaluation pillars, and efficiency statistics for UAV missions under dynamic 6G network conditions.}
\label{tab:alpha3_leaderboard}
\scriptsize
\setlength{\tabcolsep}{5pt}
\renewcommand{\arraystretch}{1.05}
\rowcolors{2}{white}{cyanblue!70} 
\begin{tabular}{lccccccccccc}
\hline
\textbf{LLM Model}
& $\boldsymbol{\alpha^3}$
& \textbf{TO}
& \textbf{SP}
& \textbf{TC}
& \textbf{IQ}
& \textbf{NR}
& \textbf{CC}
& \textbf{mean\_gen\_time\_s}
& \textbf{mean\_total\_tokens}
& $\boldsymbol{\alpha^3_{\text{per-sec}}}$
& $\boldsymbol{\alpha^3_{\text{per-1k}}}$ \\
\hline
Claude-Sonnet-4.5 & 0.949 & 1.000 & 0.965 & 1.000 & 0.961 & 0.871 & 0.492 & 62.715 & 7779.220 & 0.015 & 0.122 \\

GPT-5.2-Chat & 0.514 & 1.000 & 0.994 & 1.000 & 1.000 & 0.932 & 0.884 & 26.012 & 3913.837 & 0.020 & 0.131 \\

GPT-5-mini & 0.940 & 1.000 & 0.995 & 1.000 & 1.000 & 0.893 & 0.525 & 142.557 & 8536.894 & 0.004 & 0.069 \\

Mistral-Large-2512 & 0.888 & 1.000 & 0.885 & 1.000 & 0.971 & 0.905 & 0.563 & 95.201 & 5913.860 & 0.009 & 0.150 \\

Gemini-3-Pro-Preview & 0.174 & 1.000 & 0.952 & 1.000 & 0.987 & 0.785 & 0.697 & 119.839 & 7526.192 & 0.001 & 0.023 \\

Gemini-2.5-Flash & 0.963 & 0.977 & 0.983 & 1.000 & 1.000 & 0.890 & 0.688 & 27.090 & 5758.545 & 0.017 & 0.080 \\

GPT-4.1-mini & 0.762 & 1.000 & 0.990 & 1.000 & 0.995 & 0.895 & 0.760 & 78.300 & 4754.100 & 0.010 & 0.160 \\

ChatGPT-4o & 0.976 & 1.000 & 0.990 & 1.000 & 0.995 & 0.855 & 0.874 & 10.581 & 4032.100 & 0.092 & 0.242 \\

GPT-5.1-Chat & 0.825 & 1.000 & 0.995 & 1.000 & 1.000 & 0.917 & 0.996 & 18.994 & 2854.080 & 0.043 & 0.289 \\

Claude-Opus-4.5 & 0.818 & 1.000 & 0.980 & 1.000 & 0.944 & 0.896 & 0.539 & 57.042 & 5436.347 & 0.014 & 0.151 \\

Claude-Haiku-4.5 & 0.831 & 1.000 & 0.880 & 1.000 & 0.944 & 0.897 & 0.473 & 42.790 & 7826.720 & 0.019 & 0.106 \\

DeepSeek-V3.2-exp & 0.429 & 1.000 & 0.931 & 1.000 & 0.999 & 0.915 & 0.755 & 147.424 & 5010.350 & 0.003 & 0.086 \\

Qwen3-235B-A22B & 0.881 & 0.980 & 0.950 & 1.000 & 0.982 & 0.880 & 0.642 & 105.148 & 5652.160 & 0.008 & 0.156 \\

Mistral-Medium-3.1 & 0.715 & 0.980 & 0.964 & 1.000 & 0.996 & 0.937 & 0.469 & 115.783 & 7369.694 & 0.006 & 0.097 \\

Qwen3-Max & 0.921 & 1.000 & 1.000 & 1.000 & 1.000 & 0.918 & 0.691 & 89.623 & 5328.740 & 0.010 & 0.173 \\

Kimi-K2-Thinking & 0.502 & 1.000 & 0.989 & 1.000 & 0.963 & 0.852 & 0.621 & 314.027 & 7630.523 & 0.002 & 0.066 \\

\hline
\end{tabular}\\
\footnotesize{
$\alpha^3$ denotes the composite performance score combining six evaluation pillars.
TO (Task Outcome) measures mission completion under constraints;
SP (Safety Policy) evaluates compliance with altitude, geofencing, separation, and energy rules;
TC (Tool Consistency) measures the correctness of structured MCP and A2A interactions;
IQ (Interaction Quality) evaluates dialogue efficiency, alternation, and contextual grounding;
NR (Network Robustness) measures resilience and adaptation under degraded 6G conditions;
CC (Communication Cost) penalizes excessive token usage and tool invocations.
\texttt{mean\_gen\_time\_s} denotes the average wall-clock generation time per episode (in seconds),
\texttt{mean\_total\_tokens} denotes the average number of prompt and completion tokens per episode.
$\alpha^3_{\text{per-sec}}$ and $\alpha^3_{\text{per-1k}}$ denote efficiency-normalized composite scores per second and per thousand tokens, respectively.
}
\end{table*}

\section{Experimental Results}\label{sec:experimental_results}

The experimental setup is designed to ensure fair, reproducible, and large-scale evaluation of large language models acting as autonomous UAV agents. Each evaluation instance corresponds to the generation of a complete AI conversational UAV mission episode, where the model interacts through a multi-turn dialogue to reason, plan, and execute mission objectives. For each episode, the model receives a structured initial state describing the environment, mission goals, airspace constraints, and operational conditions, and must generate a conversational trajectory that strictly alternates between agent and user roles before terminating with a structured final state. All generated outputs must conform to a predefined UAV episode JSON schema, which is automatically enforced through post-processing and validation.

To construct a realistic, statistically significant evaluation corpus, we generate a large-scale set of 113k AI conversational UAV mission episodes based on the standardized scenarios provided by UAVBench \cite{ferrag2025uavbench}. These episodes represent complete conversational control loops in which the LLM operates as an autonomous UAV agent, making sequential decisions through dialogue-based reasoning. The corpus is built by repeatedly executing conversational UAV missions over a fixed collection of UAVBench initial states, with multiple conversational executions per scenario, resulting in a diverse yet controlled set of mission trajectories that serve as gold-standard evaluation scenarios.

For comparative benchmarking, a diverse set of state-of-the-art LLMs is evaluated, covering both proprietary and open-weight families. The candidate models include Claude (Sonnet~4.5, Opus~4.5, and Haiku~4.5), GPT models (GPT-4.1-mini, GPT-5.1-Chat, GPT-5.2-Chat, GPT-5-mini, and ChatGPT-4o-latest), Gemini models (Gemini-3-Pro-Preview and Gemini-2.5-Flash-Preview), DeepSeek variants (DeepSeek-v3.2 and DeepSeek-v3.2-exp), Qwen models (Qwen3-235B-A22B-2507 and Qwen3-Max), Mistral models (Mistral-Large-2512 and Mistral-Medium-3.1), and Kimi-K2-Thinking. This selection spans a wide range of model sizes, architectures, and optimization objectives, enabling a comprehensive comparison across quality, robustness, and efficiency.

\textcolor{black}{From the corpus of 113k AI conversational UAV mission episodes, each candidate LLM is evaluated using a fixed budget of 50 conversational episodes per UAVBench scenario. These evaluation episodes are sampled from the pre-generated corpus and reused identically across all models, ensuring that all candidates are benchmarked on the same conversational trajectories. This design balances statistical robustness with computational feasibility while preventing bias induced by scenarios or sampling.}

To control stochasticity and ensure reproducibility, generation is performed using a deterministic decoding configuration. The sampling temperature is fixed to $0.0$ for all models, and the maximum number of generated tokens per episode is capped at $10{,}000$. A global random seed of $42$ is used for experiment initialization and scenario handling. In addition, per-episode randomness is controlled using a predefined set of episode seeds $\{42, 77, 101, 2025, 1337\}$, which are consistently reused across models when generating multiple episodes per scenario. This seeding strategy enables controlled stochastic variation while ensuring identical evaluation conditions.

Episode generation uses a controlled retry mechanism to improve robustness and mitigate failures due to formatting or reasoning issues. For each conversational episode, the maximum number of generation attempts is fixed to \textbf{three}. If an attempt fails schema validation, progressively stricter constraints are appended to the prompt in subsequent attempts, explicitly restricting speaker roles and output format. The generation process terminates as soon as a valid conversational episode is produced or when all three attempts are exhausted, and the number of attempts used is recorded in the episode metadata.

In addition to reasoning quality, the setup explicitly captures computational efficiency. For each conversational episode, the wall-clock generation time is measured starting from the first attempt until a valid output is produced or all retries are exhausted. Token usage statistics reported by the inference provider are also logged, enabling the computation of total token consumption per episode and normalized efficiency metrics. All episodes are enriched with metadata including the model identifier, random seed, timestamps, generation time, number of attempts used, and detailed token usage.

Conversational episodes that fail after all retry attempts are explicitly accounted for in the evaluation. When a model fails to produce a valid conversational episode, a failure stub is written to disk rather than the episode being discarded. This design ensures that conversational generation failures are fully reflected in the evaluation statistics, enabling reliable computation of failure rates and preventing optimistic bias toward models that succeed only on simpler interactions.

\begin{figure*}[t]
    \centering
    \begin{subfigure}{0.48\textwidth}
        \centering
        \includegraphics[width=\linewidth]{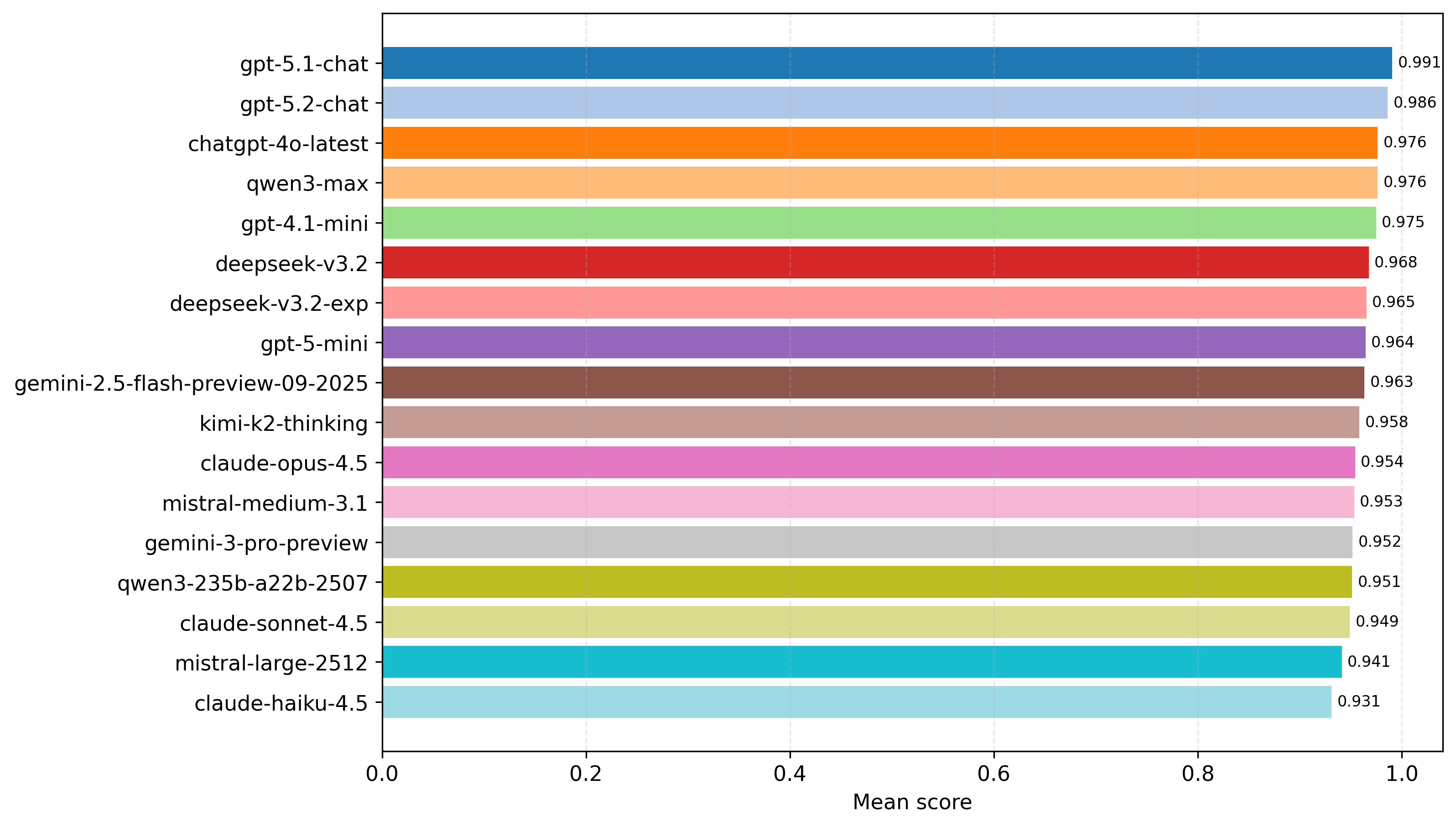}
        \caption{Mean score per LLM model.}
        \label{fig:mean-score}
    \end{subfigure}\hfill
    \begin{subfigure}{0.48\textwidth}
        \centering
        \includegraphics[width=\linewidth]{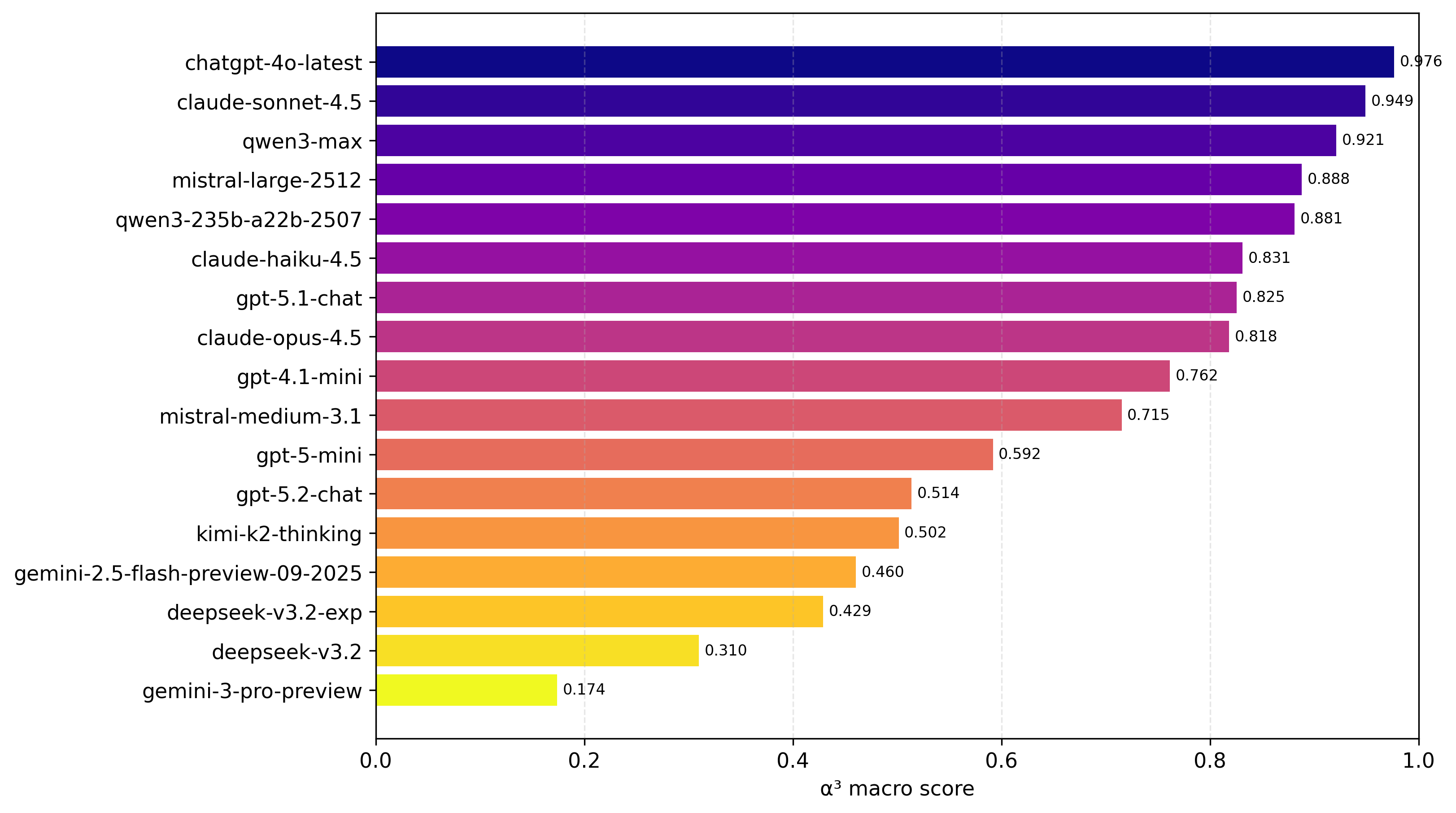}
        \caption{$\alpha^{3}$ macro score per LLM model.}
        \label{fig:alpha3-macro}
    \end{subfigure}
    \caption{Overall performance comparison of LLM agents under the $\alpha^{3}$-Bench.}
    \label{fig:overall-performance}
\end{figure*}

\subsection{Overall Performance and $\alpha^3$ Evaluation}

Table~\ref{tab:alpha3_leaderboard} presents a comprehensive comparison of state-of-the-art Large Language Models (LLMs) evaluated using the proposed $\alpha^3$-Bench framework for UAV missions under dynamic 6G network conditions. The results highlight clear differences in composite performance ($\alpha^3$), safety compliance, interaction quality, and network robustness across models. While several models achieve near-perfect Task Outcome (TO) and Tool Consistency (TC), variations in Safety Policy (SP), Network Robustness (NR), and Communication Cost (CC) significantly impact the overall $\alpha^3$ score. The table also reveals necessary trade-offs between reasoning quality and efficiency: models with high composite scores do not necessarily exhibit the best efficiency-normalized performance, as reflected by $\alpha^3_{\text{per-sec}}$ and $\alpha^3_{\text{per-1k}}$. These results demonstrate the importance of jointly evaluating mission success, safety, protocol compliance, and computational efficiency when assessing LLMs for autonomous UAV operations in realistic 6G environments. 

Figure~\ref{fig:overall-performance} summarizes the overall performance of the evaluated large language models on $\alpha^3$-Bench UAV conversational missions. The figure consists of two complementary subfigures that jointly assess average task quality and holistic efficiency-aware performance.

\paragraph{Mean Score per Model}
Figure~\ref{fig:mean-score} reports the mean task score achieved by each model across all evaluated UAV scenarios, reflecting the average correctness and coherence of model responses during mission execution. The results show that most modern LLMs achieve very high mean scores, with all models exceeding 0.93. GPT-5.1-chat achieves the highest mean score of 0.991, followed by GPT-5.2-chat at 0.986. ChatGPT-4o-latest and Qwen3-max both reach a mean score of 0.976, while GPT-4.1-mini follows closely with 0.975. DeepSeek-v3.2 (0.968), DeepSeek-v3.2-exp (0.965), and Gemini-2.5-Flash-Preview-09-2025 (0.963) also demonstrate strong performance. Even the lowest-ranked model, Claude-Haiku-4.5, maintains a mean score of 0.931. These results indicate that mean score quickly saturates for state-of-the-art models and therefore offers limited discrimination in isolation.

\paragraph{$\alpha^3$ Macro Score per Model}
Figure~\ref{fig:alpha3-macro} presents the $\alpha^3$ macro score, which integrates reasoning quality, reliability, coverage, and efficiency into a single metric. Unlike the mean score, this measure reveals substantial performance variation across models. ChatGPT-4o-latest achieves the highest $\alpha^3$ macro score of 0.976, followed by Claude-Sonnet-4.5 at 0.949 and Qwen3-max at 0.921. Mistral-Large-2512 (0.888) and Qwen3-235b-a22b-2507 (0.881) form a second performance tier. Despite achieving the highest mean score, GPT-5.1-chat records a lower $\alpha^3$ macro score of 0.825, indicating reduced efficiency or higher resource consumption. GPT-5.2-chat further drops to 0.514, while Gemini-2.5-Flash-Preview-09-2025 (0.460), DeepSeek-v3.2-exp (0.429), and DeepSeek-v3.2 (0.310) exhibit significantly weaker holistic performance. Gemini-3-Pro-Preview ranks last with an $\alpha^3$ macro score of only 0.174.

Overall, these results demonstrate that high average task accuracy does not necessarily translate into efficient or robust autonomous UAV operation. While many models achieve near-perfect mean scores, their $\alpha^3$ macro scores differ markedly, revealing essential trade-offs related to computational efficiency, coverage consistency, and failure resilience. This highlights the need for multidimensional evaluation frameworks, such as the $\alpha^3$-Bench, to assess LLM suitability for real-world autonomous UAV systems.

\begin{figure}[t]
    \centering
    \includegraphics[width=\linewidth]{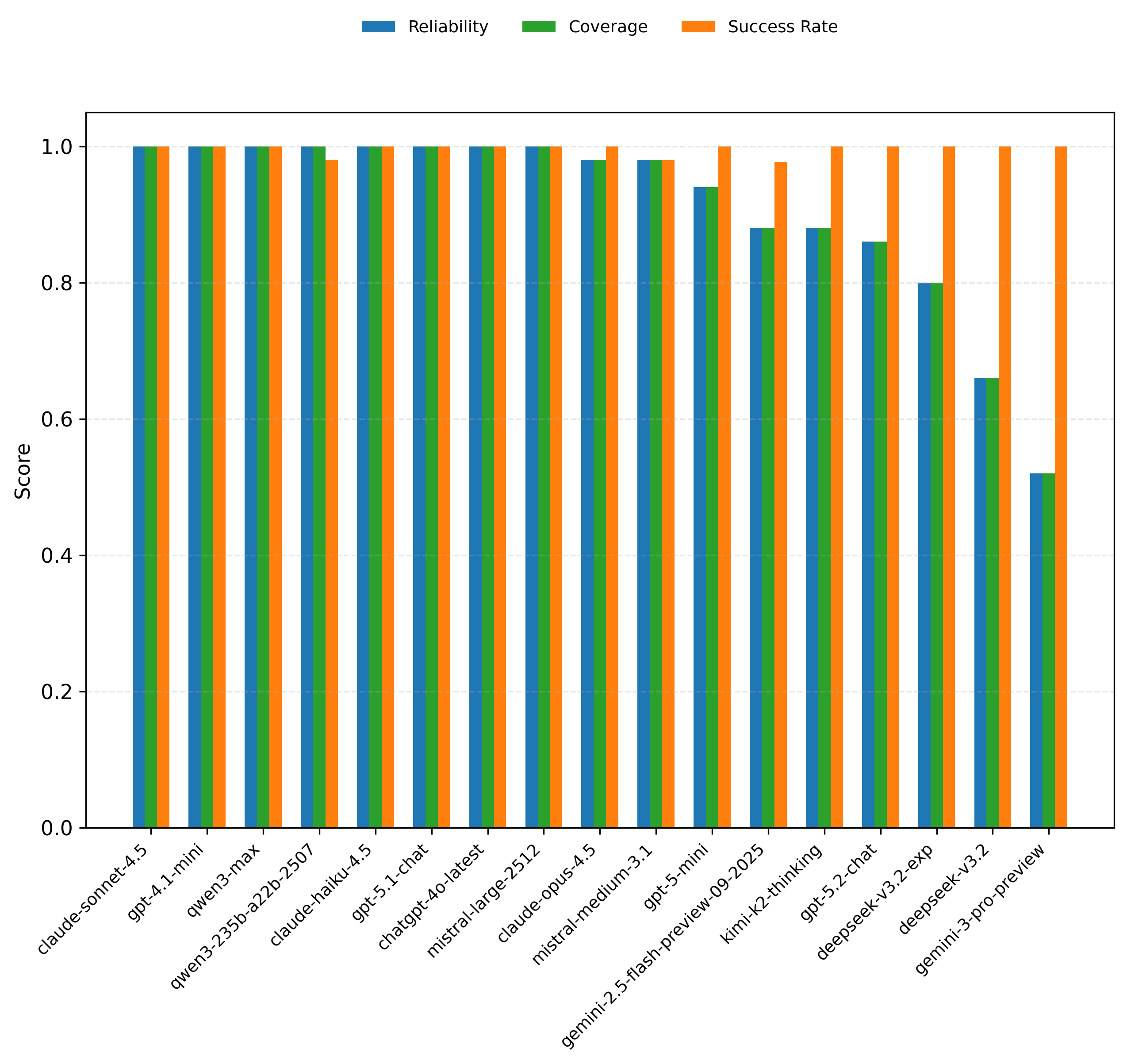}
    \caption{Reliability, coverage, and success rate across LLM models.}
    \label{fig:reliability}
\end{figure}

\subsection{Reliability, Coverage, and Failure Analysis}

\paragraph{Reliability, coverage, and success rate}
Figure~\ref{fig:reliability} illustrates the reliability, coverage, and success rate achieved by the evaluated LLM agents. Several models demonstrate near-perfect robustness across all three metrics. Claude-Sonnet-4.5, GPT-4.1-mini, and Qwen3-Max achieve reliability and coverage scores of $1.00$, together with a success rate of $1.00$, indicating fully stable mission execution. GPT-5.1-chat and ChatGPT-4o-latest similarly maintain a success rate of $1.00$ while preserving reliability above $0.99$, highlighting strong consistency under multi-turn autonomous UAV control.

A second performance tier is visible in Figure~\ref{fig:reliability} for models such as Claude-Opus-4.5 and Mistral-Medium-3.1, which achieve reliability values close to $0.98$ while still sustaining a success rate of $1.00$. In contrast, more efficiency-oriented or lightweight models exhibit a noticeable degradation. DeepSeek-v3.2 records reliability and coverage scores of approximately $0.66$, while Gemini-3-Pro-Preview drops further to $0.52$. Although these models report a nominal success rate of $1.00$, their reduced reliability and coverage indicate unstable behavior across complete mission executions.

\begin{figure}[t]
    \centering
    \includegraphics[width=\linewidth]{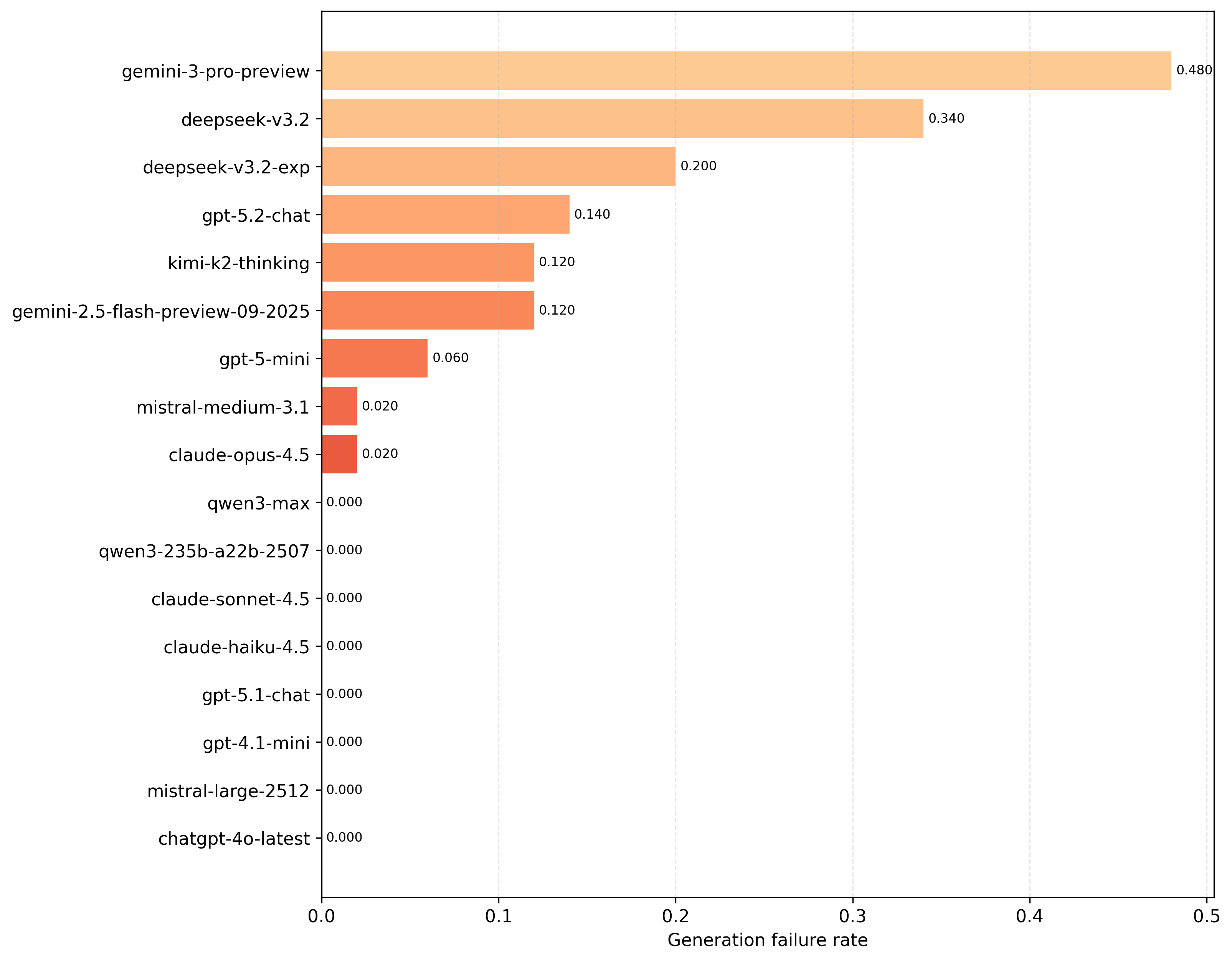}
    \caption{Generation failure rate per LLM model.}
    \label{fig:failure-rate}
\end{figure}

\paragraph{Generation failure rate}
Figure~\ref{fig:failure-rate} reports the generation failure rate for each model, offering a complementary view of execution robustness. The highest failure rate is observed for Gemini-3-Pro-Preview at $0.48$, followed by DeepSeek-v3.2 at $0.34$ and DeepSeek-v3.2-exp at $0.20$. These elevated failure rates directly explain the reduced reliability and coverage levels previously observed in Figure~\ref{fig:reliability}.

Moderate failure rates are measured for GPT-5.2-chat ($0.14$), Kimi-K2-Thinking ($0.12$), and Gemini-2.5-Flash-Preview-09-2025 ($0.12$), indicating partial instability under longer conversational trajectories. In contrast, a large subset of models—including GPT-5.1-chat, ChatGPT-4o-latest, Claude-Haiku-4.5, Claude-Sonnet-4.5, and Qwen3-235B-A22B-2507—exhibit a zero generation failure rate in Figure~\ref{fig:failure-rate}, confirming strong robustness. Overall, these findings emphasize that minimizing generation failures is essential to achieving reliable, well-covered autonomous UAV mission execution.

\begin{figure*}[t]
    \centering
    \begin{subfigure}{0.48\textwidth}
        \centering
        \includegraphics[width=\linewidth]{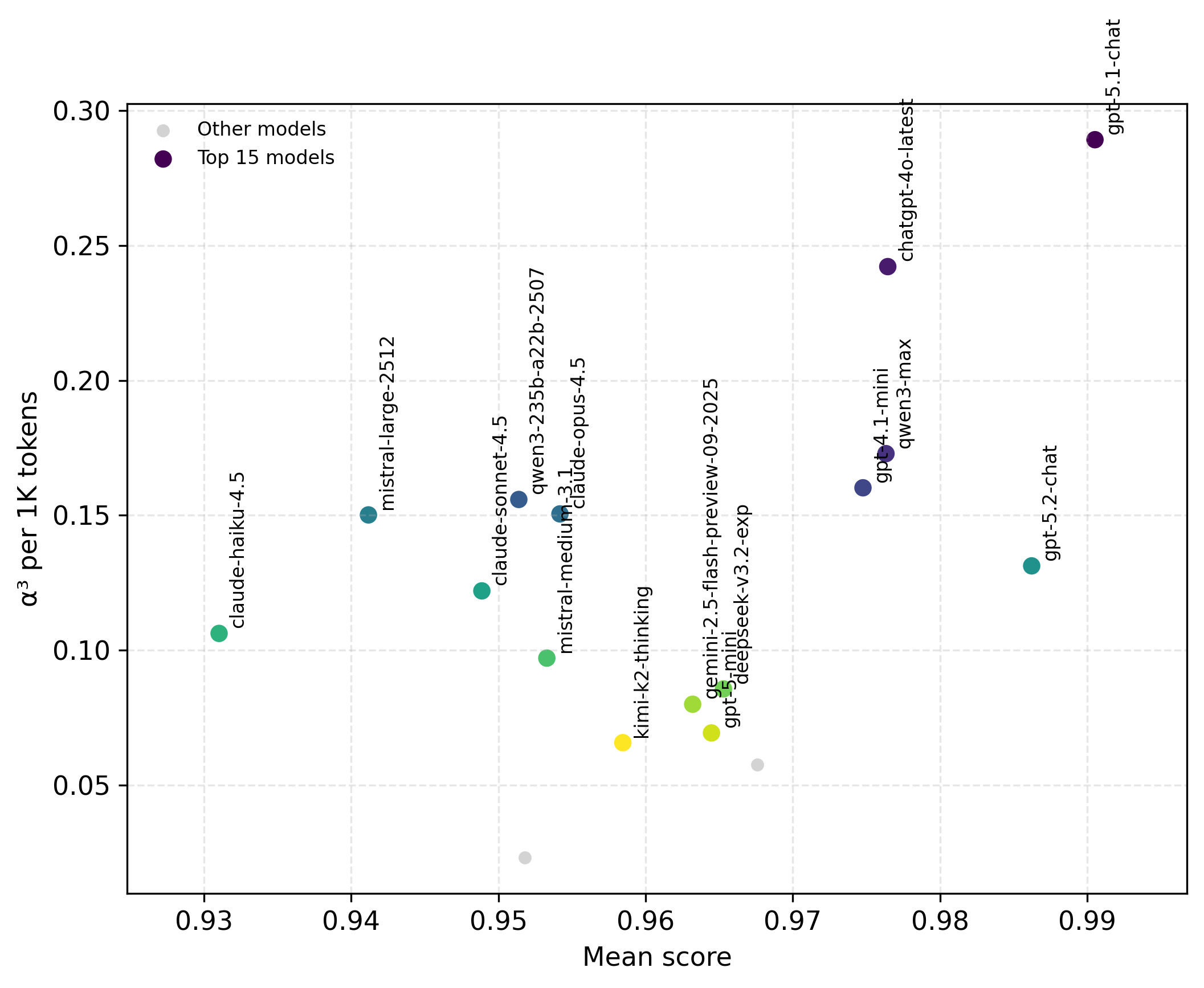}
        \caption{Quality vs. efficiency measured as $\alpha^{3}$ per 1K tokens.}
        \label{fig:efficiency-tokens}
    \end{subfigure}\hfill
    \begin{subfigure}{0.48\textwidth}
        \centering
        \includegraphics[width=\linewidth]{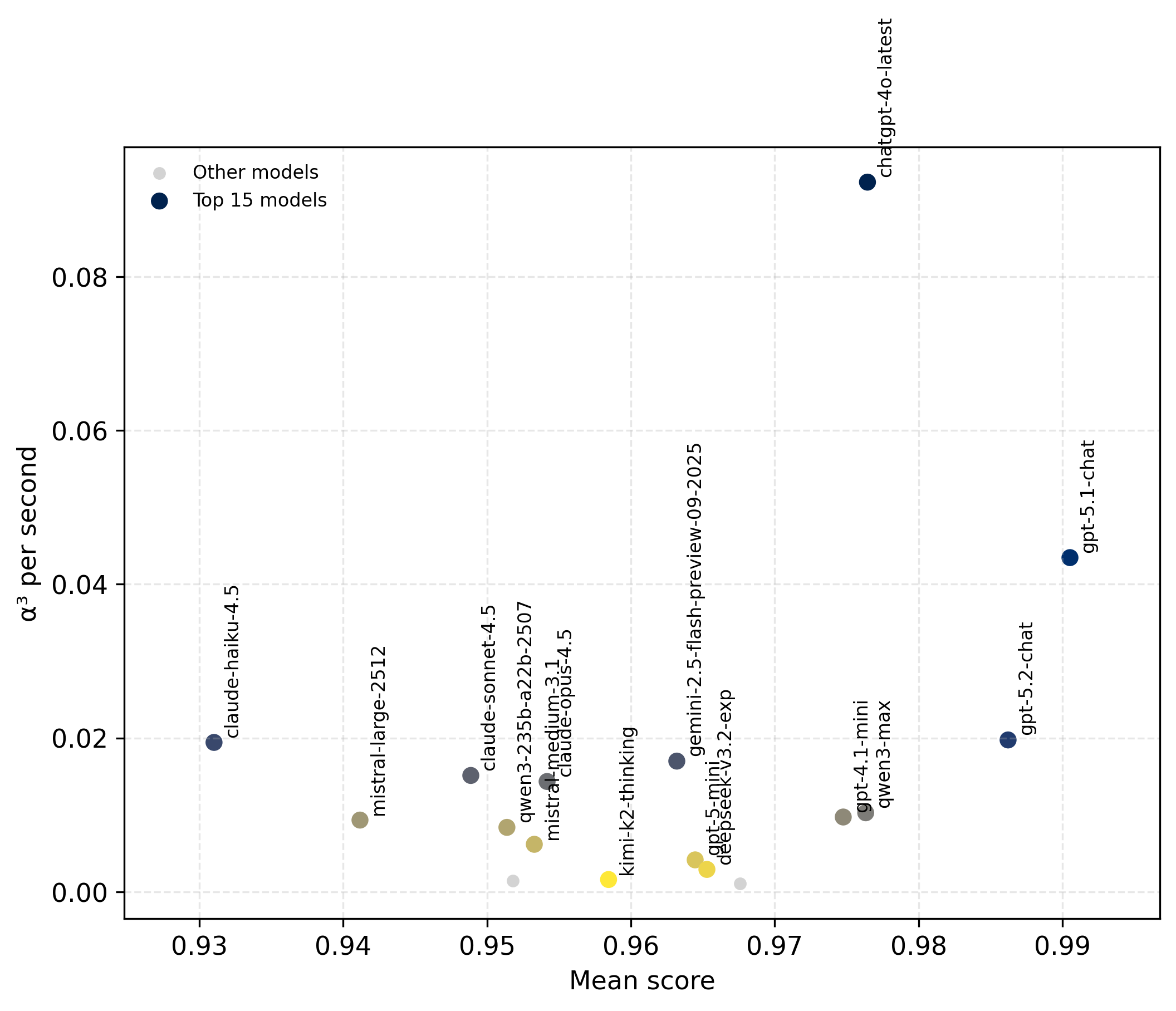}
        \caption{Quality vs. efficiency measured as $\alpha^{3}$ per second.}
        \label{fig:efficiency-time}
    \end{subfigure}
    \caption{Trade-off between reasoning quality and computational efficiency.}
    \label{fig:quality-efficiency}
\end{figure*}

\subsection{Quality–Efficiency Trade-off Analysis}

Figure~\ref{fig:efficiency-tokens} analyzes the trade-off between model quality and token efficiency by plotting the mean score against the $\alpha^3$ score per 1K tokens. Models located in the upper-right region achieve both high accuracy and strong token efficiency. The best-performing model in this setting is gpt-5.1-chat, which attains a mean score of approximately 0.99 while reaching the highest $\alpha^3$ per 1K tokens value of about 0.29. Similarly, chatgpt-4o-latest demonstrates a strong balance with a mean score close to 0.98 and a token efficiency of around 0.24. Other competitive models such as qwen3-max and gpt-4.1-mini achieve mean scores above 0.97 with $\alpha^3$ per 1K tokens values of approximately 0.17 and 0.16, respectively. In contrast, several models with comparable mean scores exhibit substantially lower efficiency. For example, gemini-3-pro-preview reaches a mean score near 0.95 but achieves only about 0.02 $\alpha^3$ per 1K tokens, highlighting a significant efficiency gap despite acceptable accuracy.

Figure~\ref{fig:efficiency-time} illustrates the relationship between mean score and temporal efficiency, measured as $\alpha^3$ per second. Once again, chatgpt-4o-latest stands out by combining a high mean score of approximately 0.98 with the highest temporal efficiency, reaching nearly 0.09 $\alpha^3$ per second. The model gpt-5.1-chat follows with a mean score close to 0.99 and a time efficiency of about 0.043. In comparison, gpt-5.2-chat and qwen3-max achieve moderate temporal efficiency values around 0.02 and 0.01, respectively, despite maintaining strong mean scores above 0.97. Several other models, including kimi-k2-thinking and deepseek-v3.2, cluster near the bottom of the plot with $\alpha^3$ per second values below 0.01, indicating that their reasoning quality comes at a substantially higher computational time cost.

Overall, Figure~\ref{fig:efficiency-tokens} and Figure~\ref{fig:efficiency-time} jointly demonstrate that high mean accuracy alone is insufficient to characterize model performance. Only a subset of models successfully translate strong reasoning capability into both token-efficient and time-efficient execution, which is essential for real-time and resource-constrained UAV mission planning.

\begin{figure*}[t]
    \centering
    \begin{subfigure}{0.48\textwidth}
        \centering
        \includegraphics[width=\linewidth]{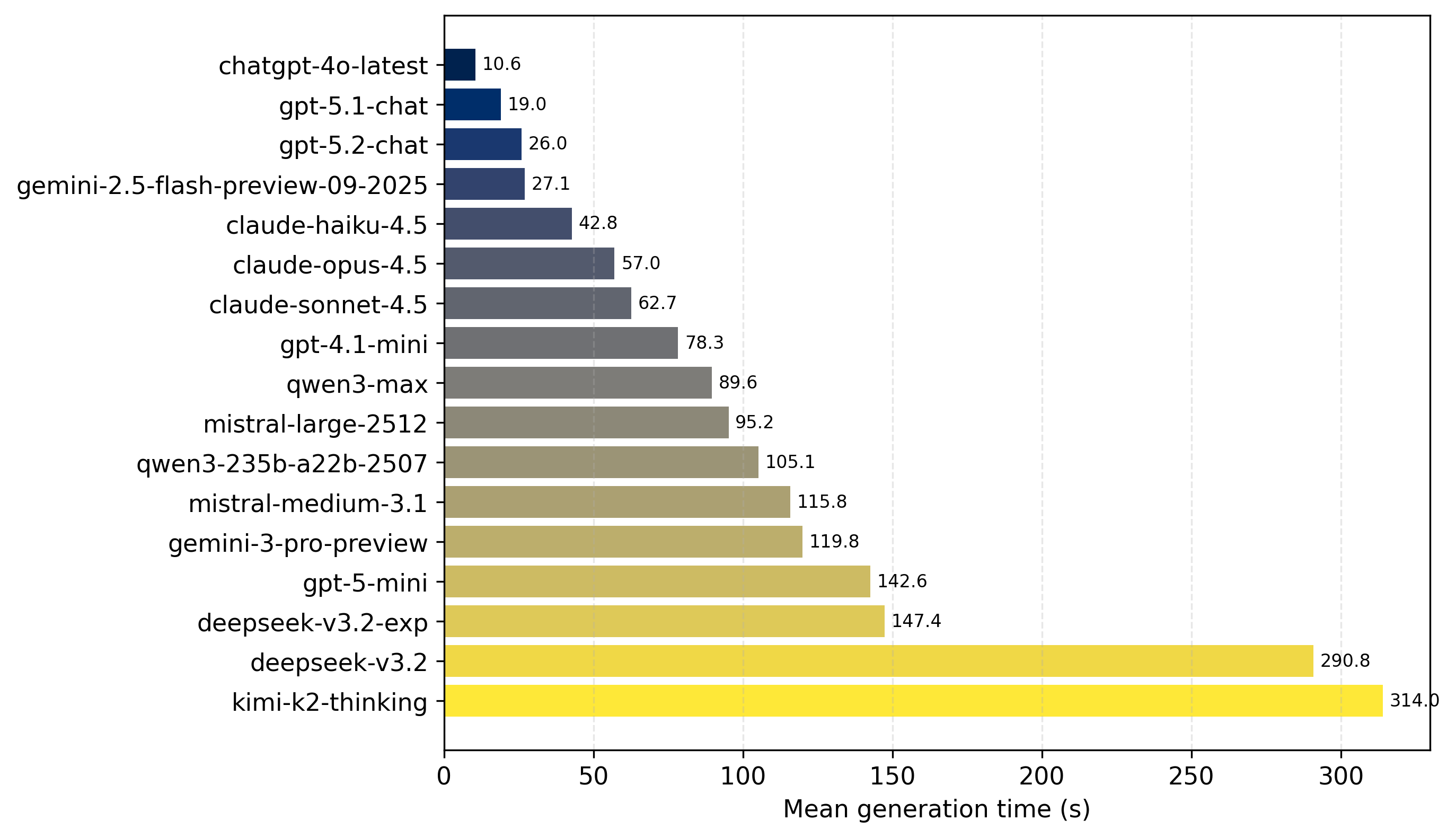}
        \caption{Mean generation time per episode.}
        \label{fig:gen-time}
    \end{subfigure}\hfill
    \begin{subfigure}{0.48\textwidth}
        \centering
        \includegraphics[width=\linewidth]{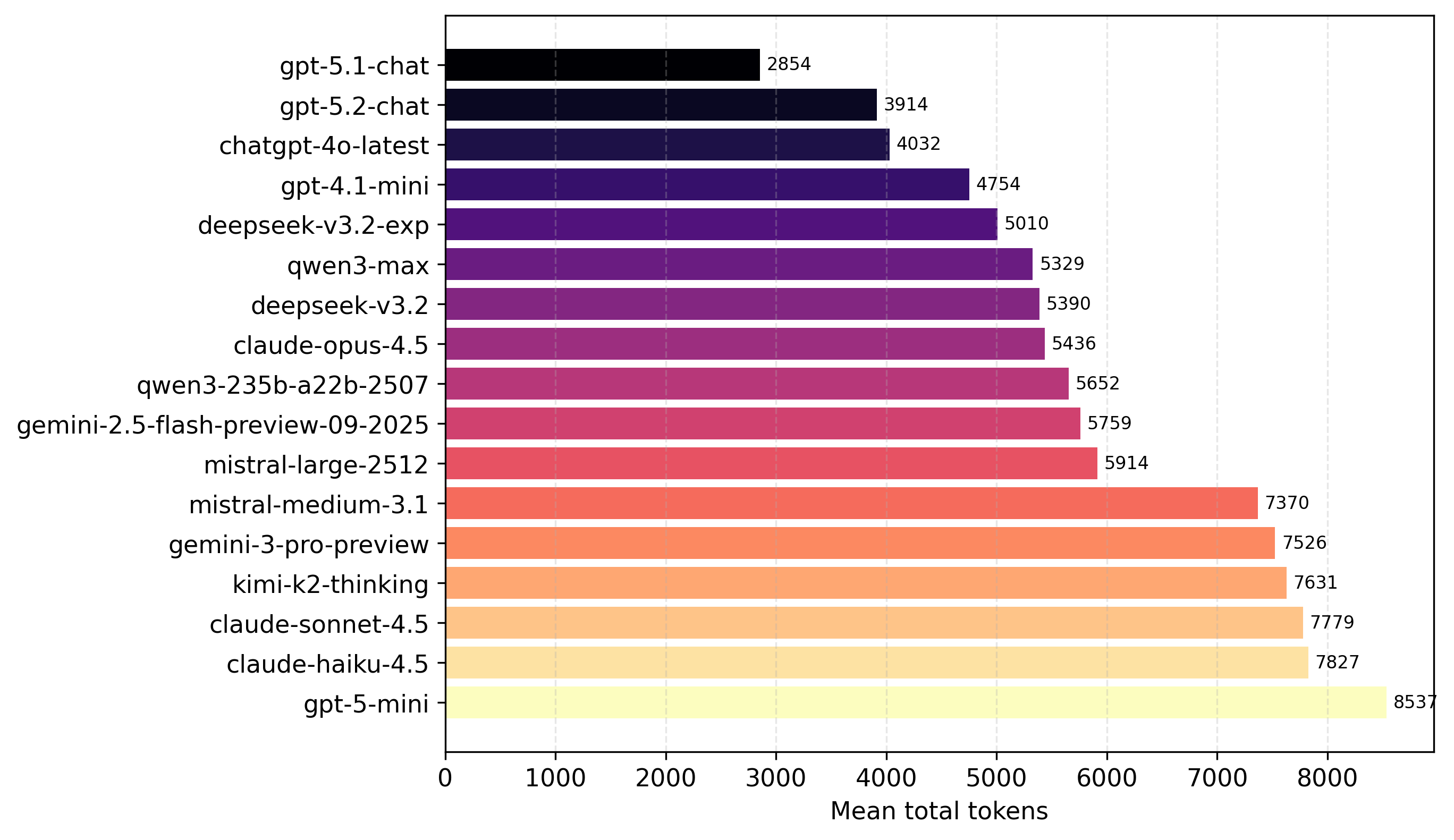}
        \caption{Mean total token consumption per episode.}
        \label{fig:token-consumption}
    \end{subfigure}
    \caption{Computational cost comparison of LLM agents.}
    \label{fig:computational-cost}
\end{figure*}

\subsection{Computational cost and resource usage}

\paragraph{Mean generation time per episode}
Figure~\ref{fig:gen-time} reports the mean generation time per episode for all evaluated LLMs, revealing substantial variation in computational latency across models. The fastest model is chatgpt-4o-latest, which completes an episode in approximately $10.6$ seconds, followed by gpt-5.1-chat and gpt-5.2-chat with $19.0$ and $26.0$ seconds, respectively. Lightweight and latency-optimized models such as gemini-2.5-flash-preview-09-2025 also exhibit low response times ($27.1$ seconds), making them suitable for time-critical UAV missions. In contrast, larger reasoning-oriented models incur significantly higher delays. For instance, mistral-medium-3.1 and gemini-3-pro-preview require $115.8$ and $119.8$ seconds per episode, respectively, while deepseek-v3.2 and kimi-k2-thinking exceed $290$ seconds, reaching up to $314.0$ seconds. This highlights a clear trade-off between advanced reasoning capabilities and real-time responsiveness in autonomous UAV operations.

\paragraph{Mean total tokens per episode}
Figure~\ref{fig:token-consumption} presents the mean total token consumption per episode, which reflects both computational and economic cost. The most token-efficient model is gpt-5.1-chat, consuming approximately $2{,}854$ tokens per episode, followed by gpt-5.2-chat ($3{,}914$ tokens) and chatgpt-4o-latest ($4{,}032$ tokens). Mid-range models such as gpt-4.1-mini and qwen3-max require between $4{,}754$ and $5{,}329$ tokens, offering a balanced trade-off between performance and cost. In contrast, reasoning-intensive models demonstrate significantly higher token usage. Notably, mistral-medium-3.1 and gemini-3-pro-preview consume $7{,}370$ and $7{,}526$ tokens, respectively, while kimi-k2-thinking reaches the highest consumption at $8{,}537$ tokens. These results indicate that while advanced reasoning improves decision quality, it introduces substantial resource overhead, potentially constraining scalability and real-time deployment in UAV systems.

\section{Conclusion}\label{sec:conclusion}

This paper addressed a fundamental and timely question: who truly wins the conversational reasoning challenge for LLM agents operating in 6G-enabled autonomous UAV systems? To answer this, we introduced $\alpha^{3}$-Bench, a comprehensive benchmark that evaluates LLM-driven UAV autonomy as a multi-turn, language-mediated control problem under realistic operational, safety, and networking constraints. Unlike prior benchmarks that focus on isolated reasoning, tool use, or perception tasks, $\alpha^{3}$-Bench captures end-to-end conversational autonomy, where an LLM must reason, communicate, and adapt continuously while respecting UAV dynamics, mission policies, and fluctuating 6G network conditions.

A key contribution of this work is the unified evaluation framework that integrates structured conversational decision-making with modern agentic protocols, namely Model Context Protocol (MCP) and Agent-to-Agent (A2A) communication. By embedding these protocols directly into each dialogue turn, the benchmark enables systematic assessment of protocol compliance, tool consistency, and multi-agent coordination alongside mission success. In addition, $\alpha^{3}$-Bench explicitly incorporates network-awareness through a dynamic 6G context vector, allowing models to be evaluated on their ability to adapt reasoning strategies under degraded latency, packet loss, and edge-computing constraints.

We constructed a large-scale corpus of 113k AI conversational UAV mission episodes grounded in UAVBench scenarios and evaluated 17 state-of-the-art LLMs using a fixed subset of 50 episodes per scenario. To enable fair and meaningful comparison, we proposed a composite $\alpha^{3}$ metric that unifies six complementary pillars—Task Outcome, Safety Policy, Tool Consistency, Interaction Quality, Network Robustness, and Communication Cost—together with reliability-adjusted and efficiency-normalized scores. Our experimental results reveal that while several frontier models achieve near-perfect mission completion and safety compliance, substantial differences remain in robustness to degraded 6G conditions, protocol adherence, generation reliability, latency, and token efficiency. These findings demonstrate that high reasoning accuracy alone is insufficient for real-world UAV deployment, and that efficiency and robustness are equally critical dimensions of conversational autonomy.

Overall, $\alpha^{3}$-Bench provides a reproducible, extensible, and failure-aware foundation for ranking LLM agents as autonomous UAV controllers rather than as isolated language models. Beyond UAV systems, the proposed formulation and metrics apply to a broader class of networked autonomous agents operating in safety-critical environments. Future work will extend $\alpha^{3}$-Bench to multimodal perception, real-world sim-to-real transfer, and learning-based policy adaptation, as well as to other 6G-enabled domains such as autonomous vehicles, robotic swarms, and edge-native AI systems. We believe that $\alpha^{3}$-Bench represents an essential step toward principled evaluation and deployment of trustworthy conversational AI agents in next-generation autonomous systems.

\bibliographystyle{IEEEtran}
\bibliography{bibliography} 

\end{document}